\documentclass[12pt]{article}
\usepackage[utf8]{inputenc}
\usepackage[english]{babel}
\usepackage{hyperref, graphicx, bm, amssymb, amsmath, epstopdf,adjustbox,doi,mathtools}
\usepackage{subfig}
\usepackage[mode=buildnew]{standalone}
\usepackage[numbers]{natbib}

\newcommand{\besK}[1][\nu]{\mathcal{K}_{#1}}
\newcommand{\besI}[1][\nu]{\mathcal{I}_{#1}}

\newcommand{\Z}{\mathbb{Z}}    
\newcommand{\R}{\mathbb{R}}    
\newcommand{\Ex}{\mathbb{E}}   
\newcommand{\bO}{\mathcal{O}}  
\newcommand{\sv}{\; | \;}      
\newcommand{\sgm}{\sigma}      
\newcommand{\bx}{\bm{x}}       
\newcommand{\bz}{\bm{z}}       
\newcommand{\bth}{\bm{\theta}} 
\newcommand{\bS}{\bm{\Sigma}}  
\newcommand{\abs}[1]{\left|#1\right|}      
\newcommand{\norm}[2][]{\left\Vert#2\right\Vert_{#1}} 
\newcommand{\set}[1]{\left\{ #1 \right\}}             
\newcommand{\matr}[1]{\begin{bmatrix*}[r]
                      #1
                     \end{bmatrix*}}  
\DeclareMathOperator*{\argmin}{arg\,min}   

\newcommand{\revision}[1]{\textcolor{black}{#1}}

\numberwithin{equation}{section}

\usepackage{tikz}
\usetikzlibrary{arrows,shapes,positioning,shadows,trees}

\renewcommand{\vec}[1]{{\mathbf{#1}}}
\newcommand{\dvec}[1]{{\dot{\vec{#1}}}}

\tikzset{                  
  level/.style   = { ultra thick, blue },
  connect/.style = { dashed, red },
  connect2/.style = { dashed, ultra thick, black },
  label/.style   = { text width=2cm }
}

\newcommand{\pkg}[1]{#1}
\newcommand{\proglang}[1]{#1}

\title{\bf Fitting Mat\'ern smoothness parameters using automatic
differentiation}
\author{
\textbf{Christopher J. Geoga} \\Department of Statistics, Rutgers University
\vspace{0.1in}\\
\textbf{Oana Marin} \\Mathematics and Computer Science Division, Argonne National Laboratory
\vspace{0.1in}\\
\textbf{Michel Schanen}\\Mathematics and Computer Science Division, Argonne National Laboratory
\vspace{0.1in}\\
\textbf{Michael L. Stein}\\Department of Statistics, Rutgers University
}
\date{}

\begin{document}

\maketitle


\bigskip
\begin{abstract}
The Mat\'ern covariance function is ubiquitous in the application of Gaussian
processes to spatial statistics and beyond. Perhaps the most important reason
for this is that the smoothness parameter $\nu$ gives complete control over the
mean-square differentiability of the process, which has significant implications
for the behavior of estimated quantities such as interpolants and forecasts.
Unfortunately, derivatives of the Mat\'ern covariance function with respect to
$\nu$ require derivatives of the modified second-kind Bessel function $\besK$
with respect to $\nu$. While closed form expressions of these derivatives do
exist, they are prohibitively difficult and expensive to compute. For this
reason, many software packages require fixing $\nu$ as opposed to estimating it,
and all existing software packages that attempt to offer the functionality of
estimating $\nu$ use finite difference estimates for $\partial_\nu \besK$. In
this work, we introduce a new implementation of $\besK$ that has been designed
to provide derivatives via automatic differentiation (AD), and whose resulting
derivatives are significantly faster and more accurate than those computed using
finite differences. We provide comprehensive testing for both speed and accuracy
and show that our AD solution can be used to build accurate Hessian matrices for
second-order maximum likelihood estimation in settings where Hessians built with
finite difference approximations completely fail.
  
\end{abstract}
\noindent%
{\it Keywords:} Mat\'ern covariance, Bessel functions, Gaussian processes, maximum likelihood, automatic differentiation
\vfill

\newpage
\section{Introduction}\label{sec:intro}

Gaussian process modeling is a ubiquitous tool in a variety of disciplines. One
attractive feature of a Gaussian process is that it is specified by its first
two moments: for a Gaussian random field $Z$ with (often multivariate) index
$\bx$, one needs only to select a mean function $\Ex Z(\bx) =:
m(\bx)$\footnote{For convenience we consider only mean-zero processes, however
extensions to parametric mean functions are straightforward.} and a
positive-definite covariance function $K$ such that $\text{Cov}(Z(\bx), Z(\bx'))
=: K(\bx, \bx')$ to completely specify the multivariate law. A classical
application of Gaussian process modeling in the mean-zero setting with data
$\bz$ is to select a parametric family of positive-definite functions
$\set{K_{\bth}}_{\bth \in \bm{\Theta}}$ and then optimize the negative
log-likelihood over the parametric family to obtain the estimator
\begin{equation} \label{eq:MLE}
  \widehat{\bth}_{\text{MLE}}
  := \argmin_{\bth \in \bm{\Theta}} - \ell(\bth \sv \bz)
   = \argmin_{\bth \in \bm{\Theta}} 
  \frac{1}{2} \left[\log\abs{\bS(\bth)} + \bz^\top \bS(\bth)^{-1} \bz \right] \ .
\end{equation}
It is common practice to then treat this MLE as a known true value and compute
subsequent estimators like interpolants or forecasts using the Gaussian law
specified by the MLE. 

One of the most popular isotropic covariance functions, the \emph{Mat\'ern}
covariance function \citep{matern1960}, is given by
\begin{equation} \label{eq:Matern}
  K_{\bth=(\sigma, \rho, \nu)} = \mathcal{M}_{\nu}(\bx, \bx') :=
  \sgm^2 \frac{2^{1-\nu}}{\Gamma(\nu)}
  \left( \frac{\sqrt{2 \nu} \norm{\bx - \bx'}}{\rho} \right)^\nu
  \besK \left( \frac{\sqrt{2 \nu} \norm{\bx - \bx'}}{\rho} \right) \ ,
\end{equation}
where $\Gamma$ is the gamma function, $\besK$ is the second-kind modified Bessel
function \citep{NIST}, and $\sigma$, $\rho$, and $\nu$ are scale, range, and
smoothness parameters respectively. The Mat\'ern class is distinguished from
other standard covariance functions by the flexibility that the  parameter $\nu$
gives, as it controls the \emph{mean-square differentiability} of the process
\citep{stein1999}. While the above functional form is complicated, the Mat\'ern
covariance is most naturally motivated by \revision{its Fourier transform pair (also
called the \emph{spectral density})}, which for $d$-dimensional processes has the
form
$
  S(\bm{f}) = \tau^2 (\zeta^2 + \norm{\bm{f}}^2)^{-\nu - d/2} \ ,
$
where $\tau$ and $\zeta$ are functions of $(\sigma, \rho, \nu)$ and $d$.  As it
turns out, $\nu$ has significant effects on the behavior of derived quantities.
Interpolants and forecasts, for example, are fundamentally different when $\nu <
1$, corresponding to a process that is not mean-square differentiable, and when
$\nu \geq 1$, corresponding to a process with at least one mean-square
derivative.  Special concern about whether processes are (mean-square)
differentiable dates back to at least Matheron \citep{matheron1989} in the
geostatistical community.  For a complete discussion on the importance of this
parameter and the value in estimating it, see \citep{stein1999}. 

A widely acknowledged challenge in GP modeling is that the log-likelihood
surface created by $\ell(\bth \sv \bz)$ is non-convex and difficult to optimize
over. Figure \ref{fig:profile_liksurf} provides an example of the log-likelihood
surface (profiled in $\sigma$, \revision{see Appendix
\ref{sec:appendix_profile}}) for a Mat\'ern process that illustrates the
non-quadratic shape of level surfaces and the difficulty of estimating both
smoothness and range parameters. A similar but arguably worse challenge occurs
for estimating the range and scale even when the smoothness parameter is known
\citep{zhang2004}.
\begin{figure}[!ht]
  \centering

\begingroup
  \makeatletter
  \providecommand\color[2][]{%
    \GenericError{(gnuplot) \space\space\space\@spaces}{%
      Package color not loaded in conjunction with
      terminal option `colourtext'%
    }{See the gnuplot documentation for explanation.%
    }{Either use 'blacktext' in gnuplot or load the package
      color.sty in LaTeX.}%
    \renewcommand\color[2][]{}%
  }%
  \providecommand\includegraphics[2][]{%
    \GenericError{(gnuplot) \space\space\space\@spaces}{%
      Package graphicx or graphics not loaded%
    }{See the gnuplot documentation for explanation.%
    }{The gnuplot epslatex terminal needs graphicx.sty or graphics.sty.}%
    \renewcommand\includegraphics[2][]{}%
  }%
  \providecommand\rotatebox[2]{#2}%
  \@ifundefined{ifGPcolor}{%
    \newif\ifGPcolor
    \GPcolortrue
  }{}%
  \@ifundefined{ifGPblacktext}{%
    \newif\ifGPblacktext
    \GPblacktexttrue
  }{}%
  \let\gplgaddtomacro\g@addto@macro
  \gdef\gplbacktext{}%
  \gdef\gplfronttext{}%
  \makeatother
  \ifGPblacktext
    \def\colorrgb#1{}%
    \def\colorgray#1{}%
  \else
    \ifGPcolor
      \def\colorrgb#1{\color[rgb]{#1}}%
      \def\colorgray#1{\color[gray]{#1}}%
      \expandafter\def\csname LTw\endcsname{\color{white}}%
      \expandafter\def\csname LTb\endcsname{\color{black}}%
      \expandafter\def\csname LTa\endcsname{\color{black}}%
      \expandafter\def\csname LT0\endcsname{\color[rgb]{1,0,0}}%
      \expandafter\def\csname LT1\endcsname{\color[rgb]{0,1,0}}%
      \expandafter\def\csname LT2\endcsname{\color[rgb]{0,0,1}}%
      \expandafter\def\csname LT3\endcsname{\color[rgb]{1,0,1}}%
      \expandafter\def\csname LT4\endcsname{\color[rgb]{0,1,1}}%
      \expandafter\def\csname LT5\endcsname{\color[rgb]{1,1,0}}%
      \expandafter\def\csname LT6\endcsname{\color[rgb]{0,0,0}}%
      \expandafter\def\csname LT7\endcsname{\color[rgb]{1,0.3,0}}%
      \expandafter\def\csname LT8\endcsname{\color[rgb]{0.5,0.5,0.5}}%
    \else
      \def\colorrgb#1{\color{black}}%
      \def\colorgray#1{\color[gray]{#1}}%
      \expandafter\def\csname LTw\endcsname{\color{white}}%
      \expandafter\def\csname LTb\endcsname{\color{black}}%
      \expandafter\def\csname LTa\endcsname{\color{black}}%
      \expandafter\def\csname LT0\endcsname{\color{black}}%
      \expandafter\def\csname LT1\endcsname{\color{black}}%
      \expandafter\def\csname LT2\endcsname{\color{black}}%
      \expandafter\def\csname LT3\endcsname{\color{black}}%
      \expandafter\def\csname LT4\endcsname{\color{black}}%
      \expandafter\def\csname LT5\endcsname{\color{black}}%
      \expandafter\def\csname LT6\endcsname{\color{black}}%
      \expandafter\def\csname LT7\endcsname{\color{black}}%
      \expandafter\def\csname LT8\endcsname{\color{black}}%
    \fi
  \fi
    \setlength{\unitlength}{0.0500bp}%
    \ifx\gptboxheight\undefined%
      \newlength{\gptboxheight}%
      \newlength{\gptboxwidth}%
      \newsavebox{\gptboxtext}%
    \fi%
    \setlength{\fboxrule}{0.5pt}%
    \setlength{\fboxsep}{1pt}%
    \definecolor{tbcol}{rgb}{1,1,1}%
    \scalebox{0.7}{
\begin{picture}(7200.00,5040.00)%
    \gplgaddtomacro\gplbacktext{%
      \csname LTb\endcsname
      \put(814,944){\makebox(0,0)[r]{\strut{}$0.5$}}%
      \put(814,1371){\makebox(0,0)[r]{\strut{}$1$}}%
      \put(814,1799){\makebox(0,0)[r]{\strut{}$1.5$}}%
      \put(814,2227){\makebox(0,0)[r]{\strut{}$2$}}%
      \put(814,2655){\makebox(0,0)[r]{\strut{}$2.5$}}%
      \put(814,3082){\makebox(0,0)[r]{\strut{}$3$}}%
      \put(814,3510){\makebox(0,0)[r]{\strut{}$3.5$}}%
      \put(814,3938){\makebox(0,0)[r]{\strut{}$4$}}%
      \put(814,4366){\makebox(0,0)[r]{\strut{}$4.5$}}%
      \put(814,4793){\makebox(0,0)[r]{\strut{}$5$}}%
      \put(981,484){\makebox(0,0){\strut{}$1.2$}}%
      \put(1779,484){\makebox(0,0){\strut{}$1.25$}}%
      \put(2578,484){\makebox(0,0){\strut{}$1.3$}}%
      \put(3377,484){\makebox(0,0){\strut{}$1.35$}}%
      \put(4176,484){\makebox(0,0){\strut{}$1.4$}}%
      \put(4975,484){\makebox(0,0){\strut{}$1.45$}}%
      \put(5773,484){\makebox(0,0){\strut{}$1.5$}}%
    }%
    \gplgaddtomacro\gplfronttext{%
      \csname LTb\endcsname
      \put(209,2761){\rotatebox{-270}{\makebox(0,0){\strut{}$\rho$}}}%
      \put(3377,154){\makebox(0,0){\strut{}$\nu$}}%
      \csname LTb\endcsname
      \put(6305,704){\makebox(0,0)[l]{\strut{}$0$}}%
      \put(6305,1389){\makebox(0,0)[l]{\strut{}$2$}}%
      \put(6305,2075){\makebox(0,0)[l]{\strut{}$4$}}%
      \put(6305,2761){\makebox(0,0)[l]{\strut{}$6$}}%
      \put(6305,3447){\makebox(0,0)[l]{\strut{}$8$}}%
      \put(6305,4133){\makebox(0,0)[l]{\strut{}$10$}}%
      \put(6305,4819){\makebox(0,0)[l]{\strut{}$12$}}%
    }%
    \gplbacktext
    \put(0,0){\includegraphics[width={360.00bp},height={252.00bp}]{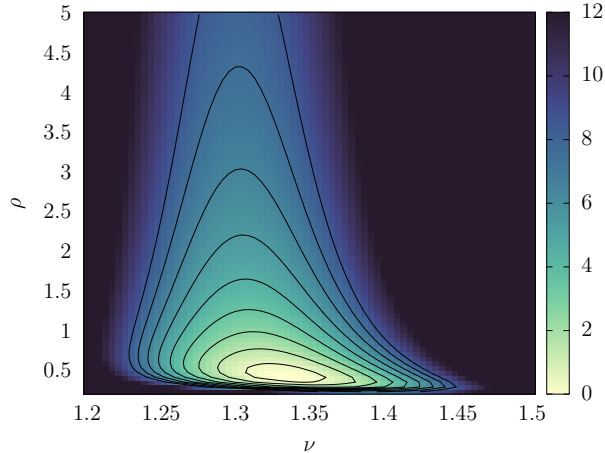}}%
    \gplfronttext
  \end{picture}%
  }
\endgroup
  \caption{A centered profile log-likelihood surface for a Mat\'ern process (so
  that $0$ corresponds to the log-likelihood at the MLE). The colorbar range has
  been artificially truncated to emphasize the non-quadratic structure near the
  minimizer. Level contours at $\tfrac{1}{2}, \tfrac{3}{2}, \ldots,
  \tfrac{17}{2}$ log-likelihood units below maximum.}
  \label{fig:profile_liksurf}
\end{figure}
An effective way to combat the log-likelihood's lack of convexity is to optimize
using derivative information. Until recently the use of derivatives to aid
Gaussian process optimization was not particularly common (see
\cite{stein2013,minden2017,geoga2020,guinness2021} for examples of separate
strategies for optimization at scale using approximated derivative information).
The gradient of $\ell(\bth \sv \bz)$ is computable in closed form and is given
by
\begin{equation} \label{eq:nllgrad}
  -2 \frac{\partial}{\partial \theta_j} \ell(\bth \sv \bz) =
  \text{tr}\left(
  \bS(\bth)^{-1} \bS_j(\bth) \right)
  -
  \bz^\top
  \bS(\bth)^{-1} \bS_j(\bth) \bS(\bth)^{-1}
  \bz \ ,
\end{equation}
where we use the shorthand $\bS_j(\bth) := \frac{\partial}{\partial \theta_j}
\bS(\bth)$.  Setting aside the computational challenges this presents for large
matrices, computing this quantity requires derivatives of the covariance
function with respect to all parameters in order to assemble the $\bS_j(\bth)$
matrices.  This is the step in the process where attempting to estimate $\nu$
from data becomes problematic. Until recently, no closed form expressions for
the derivative $\partial_\nu \besK$ were available in the literature, and by
extension accurate computation of the derivatives $\partial_\nu \mathcal{M}_\nu$
was not possible.  While these derivatives have been studied
\citep{brychkov2005}, including very recent work providing completely
closed-form derivatives \citep{gonzalez2018}, the expressions are cumbersome,
for example involving higher-order generalized hypergeometric functions, and as
such are difficult to implement reliably. \revision{Considering that these
functions are significantly more general than $\besK$, existing software
implementations for them are understandably complex and in our experience come
at significant speed costs on at least some parts of the domain, even if they
are reliably accurate on the entire domain of $\besK$.}

The non-convex optimization problem posed by maximum likelihood for Mat\'ern
covariance functions may not be practically solvable using only gradient
information. In an attempt to introduce higher-order information to further aide
in optimization, several recent works have used the expected Fisher information
matrix as a proxy for Hessians \citep{geoga2020, guinness2021}. True Hessians
(also referred to as observed information matrices), however, \revision{while
coming with an additional computational burden, are more effective than expected
information matrices in our experience as well as arguably being preferable as
proxies for second-order information of estimated parameters \citep{efron1978}. They}
can be computed as
\begin{align*} \label{eq:nllhess}
  -2[H \ell(\bth \sv \bz)]_{j,k} &=
  \text{tr} \left(
    \bS(\bth)^{-1} \bS_j(\bth)
    \bS(\bth)^{-1} \bS_k(\bth)
  \right)
  -
  \text{tr} \left(
    \bS(\bth)^{-1} \bS_{jk}(\bth)
  \right) 
   \\
  &+
  \bz^T \left(
    \frac{\partial}{\partial \theta_k}
    \set{ \bS^{-1} \bS_j(\bth) \bS^{-1} }
  \right) \bz \ ,
\end{align*}
where $\bS_{j,k}$ is the second derivative $\frac{\partial^2}{\partial \theta_j
\partial \theta_k} \bS(\bth)$.  The derivative in the last quadratic form can be
expanded using matrix calculus but is left in this form for clarity (see the
appendix of \cite{geoga2020} for the full expression).  As can be seen, the
Hessian of $\ell(\bth)$ requires second derivatives of $\besK$, making the
second-order optimization for the Mat\'ern class of covariance functions more
challenging still. \revision{To our knowledge, no existing literature gives
explicit forms for these second derivatives}.
These considerations indicate that a primary bottleneck in performing maximum
likelihood estimation for Mat\'ern models is the robust evaluation of the
second-kind modified Bessel function, $\besK$ and its first two derivatives.
\revision{While we have not directly verified that the inaccuracies of
finite-difference derivatives pose a substantial problem in gradient-based
Bayesian methods, such as Hamiltonian Monte Carlo \citep{neal2011}, it is
possible that the improvements we discuss here could also provide meaningful
gains in those methods.}

$\besK$ is a \emph{special function} defined as one of the two linearly
independent solutions of a second order differential equation \citep{NIST}.
Alternative expressions for $\besK$, such as series expansions, integrals on the
real line, path integrals, as well as confluent and generalized hypergeometric
functions have been used in existing special function libraries
\citep{NIST,amos1986algorithm,Temme1975}.  The primary difficulty of practical
implementations is that no single strategy for accurate and efficient evaluation
is valid on the entire real axis and a domain partitioning is required to obtain
an optimal method to evaluate $\besK$.  Current libraries focus entirely on
evaluations of Bessel functions, neglecting their derivatives, and even if they
provide access to the source implementation, they are not easily extended to the
problem of computing derivatives. Due to this difficulty in computing the
derivatives of $\besK$ with respect to the order $\nu$, it is common practice to
fix $\nu$ prior to performing the optimization instead of treating it as a
parameter to be estimated \revision{(see, for example, the popular
\texttt{GPyTorch} software library \citep{gardner2018}, or the
\texttt{RobustGaSP} softare \citep{gu2018,gu2018s}, both of which only offer a
few special half-integer values of $\nu$)}.  This often constitutes a
significant sacrifice, as $\nu$ can have important practical implications for
many tasks performed after maximum likelihood estimation, such as interpolation
\citep{stein1999}.

In an attempt to offer the functionality of estimating $\nu$ from data, several
software libraries for fitting Mat\'ern models provide simple finite difference
approximations for the derivative approximation as $ \partial_\nu \besK(x)
\approx h^{-1} (\besK[\nu+h](x) - \besK(x))$, where a typical choice is
$h=10^{-6}$. For special function libraries that provide only function
evaluations this is one of the few possible choices. And for a suitable choice
of $h$, these approximations may be reasonably accurate when $x$ is not near the
origin and $\nu$ is not overly large, although finite difference approximations
are known to be prone to round-off errors. This is particularly problematic for
second derivatives, as those calculations involve a divisor of $h^2$.  Combined
with the finite precision representation of the numerator, this division can
lead to severe floating point errors. To our knowledge no software library
currently attempts to offer second derivative information for $\besK$ with
respect to $\nu$, or, by extension, for $\mathcal{M}_\nu$.  Considering that the
evaluation of $\besK$ is the dominant cost in evaluating the entire Mat\'ern
covariance function, finite differences, which require at least twice as many
evaluations as direct computations, also come with a burdensome computational
cost. A more efficient approach, essentially free of round-off errors for first
derivatives, is the complex step method \citep{squire1998,fike2012automatic}.
This was studied for modified Bessel function order derivatives in
\citep{neurips21}, however not for edge cases such as $\nu\in \mathbb Z$, since
they are less problematic for complex step method evaluations but are
highly significant for the current work.  Adaptive finite difference methods,
uncommon in the statistical community, are more reliable since they select $h$
adaptively based on function values in the vicinity of differentiation points.
But while they provide higher accuracy, it comes at a significant performance
cost due to the many function evaluations required. And so while a sufficiently
high-order adaptive finite difference method can be used to solve the problem we
discuss here, the performance cost is so significant that we do not consider
them further.

Arguably the most powerful differentiation approach when derivatives are not
computable by hand or easily programmed manually is \emph{automatic
differentiation} (AD), which avoids issues such as round-off errors and allows
highly efficient implementations.  In this work, we develop a new implementation
of $\besK$ that is purposefully designed to be automatically differentiable with
high accuracy.  The negligible accumulation of errors in the first-order
derivatives using AD computations further enables highly accurate computations of
second-order derivatives of $\besK$ with respect to $\nu$.  We take this
approach here, making use of the \pkg{Julia} \citep{bezanson2017} programming
language, which we consider to be an ideal tool due to its type system
and powerful AD tools such as \pkg{ForwardDiff.jl} \citep{Revels2016}.  

In the next section, we provide an overview of AD, focusing  on the challenges
associated with an automatically differentiable implementation of $\besK$. In
Section~\ref{sec:numerics}, we present our choices for direct evaluations and
their derivatives on the entire domain of definition of $\besK$.  Next, in
Section~\ref{sec:testing}, we verify the accuracy and efficiency of the
derivatives as well as the Mat\'ern covariance. Finally, we provide a
demonstration of second-order optimization in Section~\ref{sec:demo}, where we
compare the results of optimizing using derivatives of the log-likelihood built
using our AD implementation to optimization results building the same objects
with finite difference derivatives. We conclude that
optimization using the software we provide here yields correct MLEs, and that
the AD-generated derivatives are fast and sufficiently accurate that even very
complicated derived functions such as $H \ell(\bth \sv \bz)$ are computed to
high accuracy in settings where building the same objects using finite
difference derivatives results in a complete loss of precision.

\section{Derivatives via Automatic Differentiation} \label{sec:autodiff}

\emph{Automatic differentiation} (AD) refers to the algorithmic transformation
of function implementations $f$ into code that computes the derivative $f'$.
\revision{In the forward-mode paradigm, which is the only one that will be
discussed in this work}, an implementation of $f$, representing a multivariate
vector function $\mathbf{y} = f(\vec{x}), \mathbb{R}^n \mapsto \mathbb{R}^m$
with $n$ inputs and $m$ outputs is transformed by an AD tool into an
implementation of the product $\dvec{y} = \nabla f(\vec{x}) \cdot \dvec{x} =
J_v\left(\vec{x},\dvec{x}\right)$, where $J_v$ is the Jacobian vector product of
the Jacobian $J(\vec{x})$ and the vector $\dvec{x}$. The {\it tangent vectors}
$\dvec{x}$ and $\dvec{y}$ are the input \emph{directional derivatives} and the
output sensitivities, respectively. This technique can be employed recursively
by applying AD to the Jacobian vector product implementation $\dvec{y}  =
J_v(\vec{x}, \dvec{x})$ with inputs $\vec{x}$ and $\dvec{x}$. We refer to
\citep{griewank2008} for an extensive introduction to AD.

An AD tool applies differentiation at the programming language intrinsic
function level, having hard-coded differentiation rules covering most algebraic
operations, some special functions, and optionally higher-level constructs like
linear solvers. In Julia, the programming language of choice in the current
work, these rules are implemented by \pkg{ChainRules.jl} \citep{ChainRules2021}.
The tool then generates code that computes the function value and derivative
value at each program statement.  As an example, we illustrate as pseudocode a
\revision{forward-mode} AD tool which generates $f'(x)$ from the implementation
of $y = f(x_1, x_2) = e^{(x_1 \cdot x_2)}$:
\begin{minipage}[t]{0.5\textwidth}
\begin{verbatim}
function(x1, x2)
    tmp = x1*x2
    return exp(tmp)
end
\end{verbatim}
\end{minipage}
\begin{minipage}[t]{0.5\textwidth}
\begin{verbatim}
function(x1, dx1, x2, dx2)
    tmp = x1*x2
    dtmp = dx1 * x2 + x1 * dx2
    return exp(tmp), exp(tmp)*dtmp
end
\end{verbatim}
\end{minipage}
\vspace*{5pt}

\noindent with the prefix '\verb|d|' denoting the generated tangents. Together
with the chain rule for more complex expressions, this purely mechanical work
can be fully automated.  Using compiler code analysis, like the LLVM backend of
Julia, modern AD tools provide highly efficient implementations that rival any
hand-optimized derivative code \cite{moses2020instead}.  An immediate advantage
of AD is that it enables a user to compute derivatives of a program seamlessly
without tediously differentiating by hand or contending with concerning issues
such as tuning the step size $h$ for finite differences.  The reliance of AD on
basic calculus rules implemented at the lowest computational level, as opposed
to finite differences which operate at the top level once evaluations are
available, reduces the impact of round-off errors and thus limits the
propagation of errors while being very efficiently compiled, providing even more
significant advantages for higher-order derivatives.  There are, however,
challenges associated with using automatic differentiation in the setting of
special functions, especially if applied to existing libraries, as will be
briefly discussed.

\paragraph{Differentiation of truncated series approximations} One of the most
common implementations of special functions for direct evaluations is via series
or asymptotic expansions. Term-wise derivatives of convergent series expansions
are, for the functions considered here, also convergent series, but their
convergence properties may differ from those of the original series. Thus, more
terms may be needed when calculating derivatives to assure no loss of accuracy.
Similar difficulties can occur with term-wise differentiation of asymptotic
series, whose convergence properties hold as the argument of the function tends
to some limit rather than as the number of terms increases.  From an AD
perspective a similar numerical challenge is encountered in the application of
AD to iterative methods, e.g. \cite{beck1994}. In this work, this problem occurs
primarily in one special setting that will be discussed at length in the
implementation section and is effectively remedied by adding additional terms in
the series truncation. 

\paragraph{Limit branch problems for special functions} 
AD differentiated code where the control flow is dependent on the input
$\vec{x}$ may yield discontinuous derivatives at code branches for limiting
values.  We refer to this problem as the \emph{limit branch problem}, and refer
curious readers to \citep{beck1994_if} (which calls it the \emph{if problem})
for a thorough investigation and discussion.

As an example, consider a function pertinent in this work given by
\begin{equation*} 
  f(x) := \frac{1}{2x} \left(\frac{1}{\Gamma(1-x)} -
  \frac{1}{\Gamma(1+x)}\right) \ , 
\end{equation*}
where at $x = \pm2, \; \pm3, \; ...$ one of the two gamma functions will
evaluate to infinity. A code implementation that computes $\Gamma(x)$ using a
special functions library may also throw an error at $x=0$ because there the
division of $0$ by $0$ will return \verb|NaN| per the floating point standard.
The function $f$ is analytic at $x = 0, \pm2, \; \pm3, \ldots$, however, and
admits finite limits, so any implementation of $f$ should return the correct
value for these inputs. Focusing for simplicity on the particular value of
$x=0$, one workaround is to specify the function value using a special branch
shown here as pseudocode given by
\begin{verbatim}
  function f_focus_on_0(x)
    if x == 0
      return 0.57721566... # Euler-Mascheroni constant
    else
      return (1/gamma(1-x) - 1/gamma(1+x))/(2*x)
    end
  end
\end{verbatim}
\noindent Applying AD using a tool such as \pkg{ForwardDiff.jl}
\cite{Revels2016} will incorrectly return a derivative of $0$ at $x=0$. By
manually supplying limits, the programmer has cut the dependency graph of $f$
with respect to $x$ at $x=0$, and the program reduces to a constant function
with derivative $0$.  To solve this issue, the code expression in the \verb|if|
branch has to be given in a form that preserves local derivative information in
the vicinity of $x=0$.  Local expansions in the \texttt{if} code branch allow
for the limit to be computed and differentiated numerically and ultimately
preserve derivative information.  In this example, due to the smoothness of $f$
at $x=0$ there exists a polynomial $P$ such that $f(x) \approx P(x)$ at $x
\approx 0$ (see Section~\ref{sec:int_orders}). A (pseudo)-code implementation
given by 
\begin{verbatim}
  function f_focus_on_0(x)
    if is_near(x, 0)
      return poly_eval(x, p_0, p_1, p_2, ...)
    else
      return (1/gamma(1-x) - 1/gamma(1+x))/(2*x)
    end
  end
\end{verbatim}
will now provide \revision{very accurate} derivative information at $x=0$.
\revision{This approach still of course technically gives a program
implementation with discontinuities in the derivative due to the inexact local
approximations. But if the polynomials are only substituted in a suitably small
neighborhood of the expansion points then the inexactness can be reduced to the
same order as the intrinsic inexactness of finite-precision computer arithmetic
and is thus not a concern for the present work.}

For special function implementations this concern is far from purely academic.
Many definitions of $\besK$, for example when $\nu$ is an integer, are given as
limiting forms whose implementations do not provide correct automatic
derivatives.  For this reason, applying AD transformations to existing special
function libraries without additional scrutiny is unlikely to provide correct
derivatives, which is further motivation for the current work.  As indicated in
\cite{sfnn2021}, one alternative that would sidestep these concerns would be to
use a neural network model instead of a series implementation, since automatic
differentiation is an inherent feature of neural networks, and the accuracy
could be controlled in the training phase.

\section{A Fast and Accurate Implementation of $\besK$ and $\partial_{\nu}^k\besK$}\label{sec:numerics}

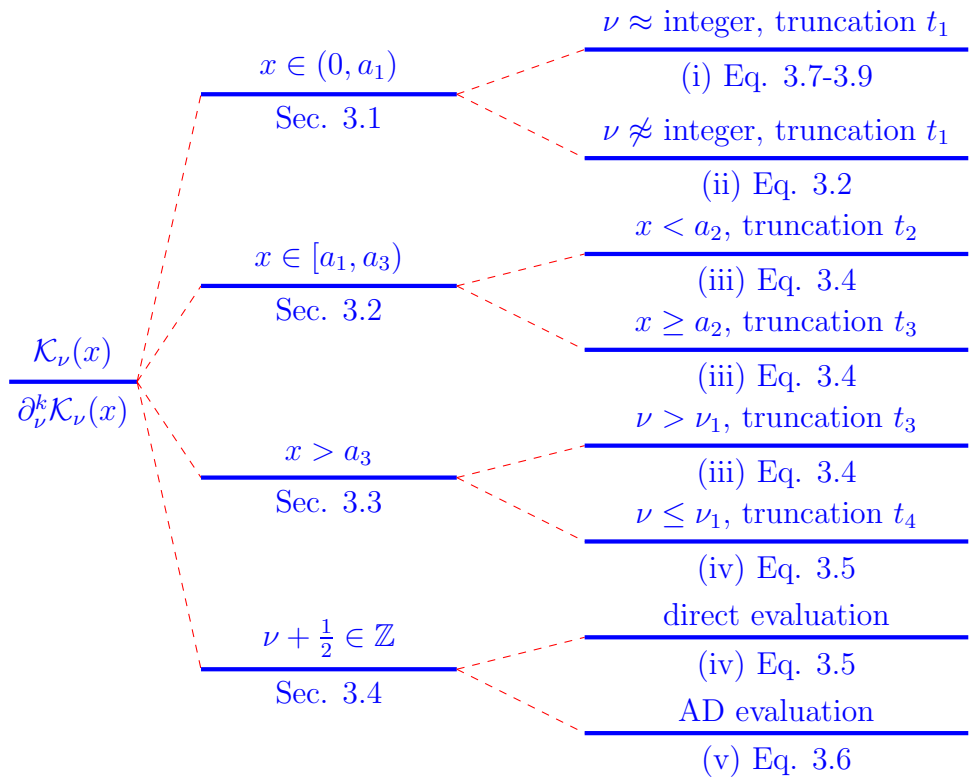
\begin{figure}[ht!]
\begin{center}
\begin{tikzpicture}[scale=0.85, auto,swap]
  \draw[level] (0,1.5) -- node[above] {$\besK(x)$}node[below] {$\partial_\nu^k \besK(x)$} (2,1.5);

  \draw[connect] (2,1.5)  -- (3,6) (2,1.5) -- (3,3) (2,1.5)--(3,0)(2,1.5)--(3,-3);

  \draw[level]   (3,6) -- node[above] {$x \in (0,a_1)$} node[below] {Sec. \ref{sec:besk_ser}} (7,6);
  \draw[level]   (3,3)  -- node[above] {$x\in [a_1,a_3)$} node[below] {Sec. \ref{sec:large_orders}} (7,3);
   \draw[level]   (3,0) -- node[above] {$x > a_3$} node[below] {Sec. \ref{sec:large_args}} (7,0);
  \draw[level]   (3,-3)  -- node[above] {$\nu+\tfrac{1}{2}\in \mathbb Z$} node[below] {Sec. \ref{sec:halfint_orders}} (7,-3);

  \draw[connect] (7,6)    -- (9,6.7);  
 \draw[connect] (7,6) -- (9,5);
 \draw[connect] (7,3)    -- (9,3.5) ;
 \draw[connect] (7,3) -- (9,2);
 \draw[connect]  (7,0) -- (9,0.5);
 \draw[connect] (7,0)    -- (9,-1) ;
  \draw[connect] (7,-3)    -- (9,-2.5) ;
 \draw[connect] (7,-3) -- (9,-4);
 
  \draw[level]   (9,6.7) -- node[above] {$\nu \approx $ integer, truncation $t_1$} node[below] {(i) Eq. \ref{eq:besk_rec}-\ref{eq:temme_ser_vp1}}(15,6.7);
  \draw[level]   (9,5) -- node[above] {$\nu \not\approx$ integer, truncation $t_1$}node[below] {(ii) Eq. \ref{eq:besk_ser}} (15,5);
    \draw[level]   (9,3.5)  -- node[above] {$x<a_2$, truncation $t_2$}node[below] {(iii) Eq. \ref{eq:besk_uae}} (15,3.5);
  \draw[level]   (9,2)  -- node[above] {$x\geq a_2$, truncation $t_3$}node[below] {(iii) Eq. \ref{eq:besk_uae}} (15,2);
   \draw[level]   (9,0.5) -- node[above] {$\nu > \nu_1$, truncation $t_3$} node[below] {(iii) Eq. \ref{eq:besk_uae}}(15,0.5);
  \draw[level]   (9,-1) -- node[above] {$\nu \leq \nu_1$, truncation $t_4$}node[below] {(iv) Eq. \ref{eq:besk_ae}} (15,-1);
  \draw[level]   (9,-2.5)  -- node[above] {direct evaluation} node[below] {(iv) Eq. \ref{eq:besk_ae}}(15,-2.5);
  \draw[level]   (9,-4)  -- node[above] {AD evaluation}node[below] {(v) Eq. \ref{eq:besk_ae_imp}} (15,-4);
\end{tikzpicture}
\caption{An overview of the branched implementation for $\besK$ and derivatives
  $\partial_\nu^k \besK(x)$. The parameters $a_i,\ t_i$ can be tuned for
  performance by the user: we recommend $a_1=8.5$, $a_2=15$, $a_3=30$,
  $t_1\approx 20$, $t_2=12$, $t_3=8$, $t_4=5$,  and $\nu_1 = 1.5$. Roman
  numerals are used by equation numbers for easy cross-referencing with Table
  \ref{tab:speedresults}. \label{fig:flowchart}}
\end{center}
\end{figure}

To devise a robust implementation for $\besK$ and derivatives
$\partial_{\nu}^k\besK$, with $k=1,2$, we partition the domain of definition,
$\R_+^2$ (this is sufficient since $\besK[\nu](x) = \besK[-\nu](x)$), into
several intervals. \revision{This strategy of breaking up the domain and
applying different strategies for different regions is ubiquitious in the
implementation of all special functions, even ones as basic as trigonometric
functions, and certainly has been applied in every commonly used implementation
of Bessel functions (see, for example, the domain partitioning of the AMOS
library \citep{amos1986algorithm}). A particular goal in our domain partitioning
and method selection, however, is to obtain accurate and efficient automatic
derivatives as well as accurate evaluations.} In Figure \ref{fig:flowchart}, we
summarize the domain partitioning based on intervals for pairs $(x,\nu)$, which
splits real axis into different sub-intervals, i.e.  $\mathbb R=(0,a_1)\cup
[a_1,a_2) \cup [a_2,a_3) \cup[a_3,\infty)$, with recommended selections for the
parameters $a_i,\ i=1,2,3$. \revision{Even in the most rigorous settings of
error analysis, however, there is an element of personal judgment in the exact
choices of domain partitioning parameters. While in broad strokes they are
guided by when different representations that can be exactly evaluated by a
computer are accurate, such as large-argument expansions being picked when the
argument is ``large enough", they are also guided by practical metrics like
numerical tests. This is the case in this work, and as such the boundary values
we recommend here can be altered slightly without major impacts on accuracy.}

Each interval is studied in its associated subsection, accompanied by
discussions on implementation details \revision{for the selected method in that
interval} (typically a suitable series-type representation) and adjustments
necessary to expedite and preserve the accuracy of automatic derivatives. Each
separate expansion is truncated at different levels $\{t_j\}$, and these
parameters have a meaningful impact on both efficiency and accuracy. In the
present work we chose the $\{t_j\}$ levels that provide close to computer
precision.  A user willing to sacrifice a few digits of accuracy to increase the
efficiency may prefer to truncate the series at lower levels than the suggested
values, but a comprehensive study of accuracy and efficiency impacts at such
levels may be lengthy and is not considered in this work.  Special values of the
order $\nu$, such as integers and half-integers must be be considered
separately. Since $\besK$ for these values of $\nu$ is commonly expressed in a
limiting form, substantial work is required to develop an implementation that
avoids the limit branch problem. As the rescaled function $x^\nu \besK(x)$
appears in the Mat\'ern covariance function, we also provide discussion of
special optimizations for this function in Section \ref{sec:rescale} of the
appendix.

\subsection{Small arguments and small-to-moderate orders} \label{sec:besk_ser}

While truncated direct series implementations may pose numerical difficulties in
certain regimes, they can nonetheless be fast and accurate when applicable.
Here, we identify a numerically stable series representation that extends the
small argument interval up to $x\approx 8.5$ while controlling the accumulation
of round-off errors at satisfactory levels.  Consider the definition of $\besK$,
for non-integer orders $\nu$, given in terms of the modified Bessel function of
the first kind, $\besI$ as
\begin{equation} \label{eq:besk_besi}
\besK(x) = \frac{
  \pi (\besI[-\nu](x) - \besI(x))
}{
  2 \sin \pi \nu
}
, \quad \text{with} \quad
  \besI(x) = (\tfrac{1}{2} x)^\nu \sum_{k=0}^\infty
  \frac{
    \left(x^2/4 \right)^k
  }{
    k! \Gamma(\nu + k + 1)
  } \ .
\end{equation}
The function $\besI$ exhibits fast growth (asymptotically on the order of
$e^x/\sqrt{2 \pi x}$); thus for large arguments Equation~\ref{eq:besk_besi}
leads to significant floating-point errors. Although the expression is valid for
all arguments, it has practical use only for sufficiently small $x$.  The
threshold for what constitutes ``sufficiently small" is chosen based on accuracy
and efficiency considerations, and in \cite{Temme1975}, values in the vicinity of
$x=4$ are chosen as the cutoff.

\revision{ An improved series expansion, however, can be derived using basic
properties of the Gamma function $\Gamma(x)$. Using Euler's reflection property
(which states that $\sin \pi x = \pi (\Gamma(1-x) \Gamma(x))^{-1}$) and
regrouping terms conveniently, we obtain a particularly convenient form for the
series expansion of $\besK$ given by}
\begin{equation} \label{eq:besk_ser}
  \besK(x) =
  \sum_{k=0}^\infty
  \left( \frac{x}{2} \right)^{2k}   \frac{1}{2k!}
  \left(\Gamma(\nu) \left( \frac{x}{2} \right)^{-\nu}
  \frac{\Gamma(1-\nu)}{\Gamma(1+k-\nu)} +\Gamma(-\nu) \left(\frac{x}{2} \right)^\nu
  \frac{\Gamma(1+\nu)}{\Gamma(1+k+\nu)}	\right) \ .
\end{equation}
One advantage that the series in Equation \ref{eq:besk_ser} has over the
expression in Equation \ref{eq:besk_besi} is that it is less prone to
cancellation errors by avoiding the subtraction of terms that grow rapidly with
the series index or argument $x$.  The series as stated in Equation
\ref{eq:besk_ser} is infinite; however a computer implementation of the infinite
series in \ref{eq:besk_ser} requires a truncation level or stopping criterion.
To promote accuracy we chose to stop the accumulation of terms when a tolerance
of $\epsilon \leq 10^{-12}$ is met, which in practice corresponds to
approximately $20$ or fewer terms in the sum. For consistency across all
computational branches we specify in Figure~\ref{fig:flowchart} a truncation at
$t_1\approx 20$, while reminding the user that in practice the stopping
criterion is tolerance-based and can be adjusted.  Furthermore, using the
property $\Gamma(x+1) = x\Gamma(x)$, it is possible to evaluate this series (in
fact its numerical truncation) using entirely algebraic operations after only
evaluations of $\Gamma(\nu)$ and $\Gamma(-\nu)$. Applying AD to Equation
\ref{eq:besk_ser} with the same stopping criterion as for the direct evaluation
results in convergence in approximately as many terms for both first and second
order derivatives. Although we do not derive a formal convergence proof, we will
show in Section~\ref{sec:testing} that the performance of this expression is one
of the best across the entire domain of definition.

For integer orders, the $\sin \pi \nu$ term in the denominator of
Equation~\ref{eq:besk_besi} leads to a singularity, and the expression is no
longer valid in such cases. The function is defined at these orders, however,
and by taking appropriate limits, $\besK$ can be written at integer orders as
\begin{equation} \label{eq:besk_intv}
2 (-1)^{n-1} \besK[n](x) := \left(
\frac{\partial \besI[\nu](x)}{\partial \nu} \bigg|_{\nu=n}
+
\frac{\partial \besI[\nu](x)}{\partial \nu} \bigg|_{\nu=-n}
\right) \ .
\end{equation}
While such integer-order expressions are valid in the limit and may be
convenient to evaluate, they lead to limit branch problems that cause incorrect
automatic derivatives, as described in Section~\ref{sec:autodiff}.  Therefore
the integer case will be considered separately in Section~\ref{sec:int_orders}.
Additionally, we comment in the Appendix on implementation specializations to
compute $x^\nu \besK(x)$ directly, which in some cases can improve speed and
accuracy.

\subsection{Intermediate arguments} \label{sec:large_orders}

Asymptotic expansions resemble series expansions in that they are also
cumulative sums which require similar implementations, but they may diverge
outside a prescribed range.  Within the range of large orders $\nu$ and
intermediate arguments $x\in[a_1, a_2)$ we introduce the \emph{uniform
asymptotic expansion} (UAE) in $\nu$ for $\besK$.  As indicated by the word
uniform, this approximation is not highly oscillatory in the argument $x$, and
as such it can be applied in argument ranges where round-off errors can be
problematic.  The UAE is given by
\begin{equation} \label{eq:besk_uae}
  \besK(\nu x) \mathop{\sim}_{\nu \rightarrow \infty}
  \sqrt{\frac{\pi}{2 \nu}}
  \frac{
    e^{-\nu \revision{\eta(x)}}
  }{
    (1 + x^2)^{1/4}
  }
  \sum_{k=0}^{\infty}
  (-1)^k \nu^{-k} U_k(\revision{p(x)})
\end{equation}
where, as in \citep[Eq.~10.41.4]{NIST}, we have
\begin{align*} 
  \eta(x) := \sqrt{1+x^2} + \log \left( \frac{x}{1 + \sqrt{1+x^2}}  \right), \quad   p(x)    := (1+ x^2)^{-1/2} \ ,
\end{align*}
and the polynomials $U_k$ are given by $U_0(p) \equiv 1$ and
\begin{equation*} 
  U_{k+1}(p) = \frac{1}{2} p^2 (1 - p^2) U_k'(p) + \frac{1}{8} \int_1^p (1 - 5t^2)
  U_k(t) \text{d} t \ .
\end{equation*}
Tables of the coefficients for the $U_k$ polynomials exist (see
\cite[Sec.~10.41]{NIST}), but they may be difficult to find digitally. Since
they are not difficult to compute using symbolic operations, our software
computes them directly (but ahead of time, to be re-used) such that the
truncation order of the approximation can be arbitrarily chosen by the user.

For reasonably large truncation orders (which in our case default to $t_2=12$ or
$t_3=8$ depending on the magnitude of $x$), this approximation is very fast to
compute. As briefly noted in \cite[Sec.~10.41]{NIST}, due to the uniqueness
property of asymptotic expansions, this expansion must agree with the more
commonly encountered expansion for large arguments to be introduced in Section
\ref{sec:large_args}. From this perspective, it comes as little surprise that
the approximation quality of Equation \ref{eq:besk_uae} for any fixed truncation
order improves as the argument \emph{or} order increases, although not
necessarily at the same rate \revision{(see \citep[10.41(iv)]{NIST} for some
discussion and a precise error bound when viewing (\ref{eq:besk_uae}) as a
generalized asymptotic expansion in $x$)}. This strategy for evaluating $\besK$
is preferable in cases where small-argument methods start to lose accuracy but
where the large-argument asymptotic expansion is not yet of satisfactory
accuracy (for example, $x = 10$ and $\nu = 3$). Moreover, since this expression
reduces to polynomial computations, it is highly stable under automatic
differentiation and less prone to round-off errors. 

\subsection{Large arguments} \label{sec:large_args}

For large arguments it is preferable to use a reduced asymptotic expansion,
equipped with precise error bounds derived in \cite{nemes2017} (\revision{see
the next section for more explicit details}), given as in
\citep[Eq.~10.40.2]{NIST} by
\begin{equation} \label{eq:besk_ae}
  \besK(x) \mathop{\sim}_{x \rightarrow \infty} 
  \revision{\sqrt{\frac{\pi}{2x}}} e^{-x}
  \sum_{k=0}^{\infty} x^{-k} a_k(\nu) \ ,
\end{equation}
 where $a_k(\nu)$ is generated via $a_0(\nu) \equiv 1$ and
\begin{align*} 
 a_k(\nu) :&=
  \frac{
    (4 \nu^2 - 1^2)  (4 \nu^2 - 3^2) \cdots (4 \nu^2 - (2k-1)^2)
  }{
    8^k \Gamma(k+1)
  } \ .
\end{align*}
For arguments $x > 30$ and moderate truncation orders such as $t_4=5$ (the
default in our implementation), this approximation is accurate to computer
precision \revision{(see \citep[10.40(iii)]{NIST} for details on the error term,
which without the exponential improvement introduced in the next section is
challenging to work with)}. The automatically differentiated implementation is
extremely fast since it involves differentiation of algebraic expressions, thus
free of numerical artifacts, and is applied to a very low number of terms up to
$t_4=5$, which leads to a very small operation count.

\subsection{Half-integer orders} \label{sec:halfint_orders}

An interesting property of $\besK$ expressed via the asymptotic expansion in
Equation \ref{eq:besk_ae} is that, when $\nu+1/2 \in \Z$, terms with $k > \nu$
drop out, and thus (\ref{eq:besk_ae}) is exact for all $x$. These cases reduce
to simpler forms that can be manually included in a program (for example,
returning the simplified $\besK[1/2](x) = \sqrt{\pi/(2x)}e^{-x}$ for $\nu =
1/2$), but this type of program implementation would cause limit branch problems
with respect to $\nu$ (as discussed in Section $2$), and as such requires
additional scrutiny in our setting to ensure accurate automatic derivatives with
respect to order. 

As noted in Section~\ref{sec:autodiff}, a converged direct evaluation may not
yield a differentiated series that has converged to the same degree.  This can
be remedied using what \cite[Sec.~10.40(iv)]{NIST} refers to as
\emph{exponentially-improved} asymptotic expansions. The asymptotic expansion
introduced in Equation
\ref{eq:besk_uae} can be re-written as
\begin{equation} \label{eq:besk_ae_imp}
  \besK(x) = \sqrt{\frac{\pi}{2x}} e^{-x} \left(
    \sum_{k=0}^{l-1} x^{-k} a_k(\nu) + R_l(\nu, x)
  \right) \ ,
\end{equation}
where the remainder term $R_l(\nu, x)$, \revision{itself reasonably complicated
to bound neatly (see \citep[10.40(iii)]{NIST})}, can then be expanded as
\begin{equation*} 
  R_l(\nu, x) = (-1)^l 2 \cos(\nu \pi)
  \left(
    \sum_{k=0}^{m-1} x^{-k} a_k(\nu) G_{l-k}(2x) + R_{m,l}(\nu, x)
  \right) \ .
\end{equation*}
Here 
$
  G_{p}(x) = \frac{e^x}{2 \pi} \Gamma(p) \Gamma(1-p, x),
$
where $\Gamma(s,x)$ is the upper incomplete gamma function
\citep[Eq.~10.17.16]{NIST}, \revision{and $R_{m,l}$ is a new and even more
finely controlled remainder term that admits a much simpler controlling bound of
$\bO(e^{-2x} x^{-m})$ \cite{nemes2017, NIST}}. While direct evaluations of
$\besK$ in the case where $\nu + \tfrac{1}{2} \in \Z$ are exact, the remainder
term is required to preserve derivative information, \revision{as the above
additional expansion makes clear that} $\partial_{\nu}^kR_{l}(v,x)$ is not zero
for any $k$. \revision{As such, additional terms from the $R_l$ approximation
are necessary to preserve accurate derivative information.}

\subsection{Integer orders and small arguments with large orders} \label{sec:int_orders}

A common strategy for computer implementations of $\besK$ is to develop an
implementation for $\nu$ in a neighborhood of $0$ and use a standard recursion
with respect to order, valid for $\besK[\nu]$ at any $\nu\in\mathbb R$, given by
\begin{equation} \label{eq:besk_rec}
  \besK[\nu+1](x) = \frac{2\nu}{x} \besK(x) - \besK[\nu-1](x) \ .
\end{equation}
For integer orders, several options are available to compute the initial
recursion terms $\besK[0]$ and $\besK[1]$, for example direct polynomial or
rational function approximations for $\besK[0]$ and $\besK[1]$ (as provided by
\citep{abramowitz65}). However, this strategy will not be helpful in providing
$\partial_\nu^k \besK$ for $\nu \in \Z$ due to the limit branch problem at $\nu =
0$ and $\nu = 1$.

A suitable approach, provided in \cite{Temme1975}, is to consider series
expansions of the recursion terms $\besK$ and $\besK[\nu+1]$, which due to
(\ref{eq:besk_rec}) (and the property that $\besK(x) = \besK[-\nu](x)$)
need only be considered on the interval $\nu \in [-1/2, 1/2]$. Reproducing the
notation of \cite{Temme1975}, we consider the expressions
\begin{align} 
  \besK(x)        &= \sum_{j=0}^{\infty} c_j f_j(\nu, x) \quad \text{and}
  \label{eq:temme_ser_v} \\
  \besK[\nu+1](x) &= \frac{2}{x} \sum_{j=0}^{\infty} c_j (p_j(\nu, x) - jf_j(\nu, x)) \ ,
  \label{eq:temme_ser_vp1} 
\end{align}
where for  $\mu(x) = \nu \log(2/x)$ we have
\begin{align}\label{eq:f0first} 
  f_0(\nu, x) &= \frac{\nu \pi}{\sin \pi \nu} \left[\Gamma_1(\nu) \cosh \mu(x)
  +
  \Gamma_2(\nu) \log(2/x) \sinh(\mu(x))/\mu(x) \right], \\ \nonumber
  \Gamma_1(\nu) &= \left(\Gamma(1-\nu)^{-1} - \Gamma(1+\nu)^{-1}\right)/(2\nu), \quad
  \Gamma_2(\nu) = \left(\Gamma(1-\nu)^{-1} + \Gamma(1+\nu)^{-1}\right)/2,  \nonumber
\end{align}
and all subsequent $f_k$ terms can be computed via a direct recursion using
\begin{align*} \label{eq:}
  p_0(\nu, x) &= \frac{1}{2} \left( \frac{x}{2} \right)^{-\nu} \Gamma(1 + \nu),
  \quad  p_j = p_{j-1}/(j-\nu) \ ,  \\
  q_0(\nu, x)  &= \frac{1}{2} \left( \frac{x}{2} \right)^{\nu} \Gamma(1 - \nu),
  \quad  q_j = q_{j-1}/(j+\nu) \ , \quad
  \text{and} \\
  f_j &= (j f_{j-1} + p_{j-1} q_{k-1})/(j^2 - \nu^2) \ .
\end{align*}
With these expressions, the problem of obtaining an automatically differentiable
program for $\besK[n](x)$ for all $n$ is reduced to obtaining an AD-compatible
implementation of $f_0(\nu,x)$ for $\nu$ in a vicinity of the origin.  As
indicated in Section \ref{sec:autodiff}, the limit branch problem can be
resolved using Taylor expansions in Equation~\ref{eq:f0first} about $\nu = 0$: 
\begin{eqnarray*} 
  \Gamma_1(\nu) &\approx &
  1 + \frac{\gamma^2 - \pi^2/6}{2} \nu^2 +
  \frac{60 \gamma^4 - 60 \gamma^2 \pi^2 + \pi^4 - 240 \gamma
  \psi^{(2)}(1)}{1440}\nu^4 \ , \\
  \Gamma_2(\nu) &\approx &
  \gamma + \frac{2 \gamma^3 - \gamma \pi^2 - 2 \psi^{(2)}(1)}{12} \nu^2
  + \\
  &&+\frac{12 \gamma^5 - 20 \gamma^3 \pi^2 + \gamma \pi^4 - 120 \gamma^2 \psi^{(2)}(1)
  + 20 \pi^2 \psi^{(2)}(1) - 12 \psi^{(4)}(1)}{1440} \nu^4 \ ,
  \\
  \frac{\sinh \mu}{\mu} &\approx & 1 + \mu^2/6 + \mu^4/120, \quad \text{and} \quad 
  \frac{\pi \nu}{\sin \pi \nu} \approx  1 + (\pi \nu)^2/6 + 7(\pi \nu)^4/360 \ .
\end{eqnarray*}
With this strategy, we obtain an expression for $f_0$ which is exact at $\nu =
0$ and very accurate for $\nu \approx 0$.  By precomputing the coefficients of
Taylor expansions, this entire apparatus yields function evaluations that are
only approximately twice the cost of the direct series that we introduce above,
and whose AD performance is quite satisfactory both with respect to speed and
accuracy. For specific details on this recursion, such as accurate and efficient
evaluations of $\Gamma_1$ and $\Gamma_2$ using Chebyshev expansions, we refer to
\citep{Temme1975}. As with Section \ref{sec:besk_ser}, we additionally provide
in the appendix a discussion of extending this method to the direct computation
of $x^\nu \besK(x)$ for slight additional gains in accuracy and speed.

\section{Speed and Accuracy Diagnostics}\label{sec:testing}

This section will focus on the verification and efficiency of our implementation
of $\besK$ and its derivatives, which we will denote either $\partial_\nu^1
\besK$ or $\partial_\nu^2 \besK$.  We compare our implementation, denoted
$\tilde{\besK}$, with the one provided by the \pkg{AMOS} library
\citep{amos1986algorithm}, $\besK^A$, using a reference solution $\besK^R$,
considered to be the ground truth and computed using the arbitrary-precision
\pkg{Arb} library \citep{Johansson2017}.  Derivatives of up to second order
computed in two ways will be compared, namely AD-generated derivatives of our
implementation, $\partial_\nu^{k\text{AD}} \tilde{\besK}$, and non-adaptive
finite difference-based derivatives of the AMOS library,
$\partial_\nu^{k\text{FD}} \besK$. These two values will be compared against a
reference derivative computed using tenth-order adaptive finite difference
methods of $\besK^A$, denoted as $\partial_\nu^{k\text{R}} \revision{\besK}$,
computed using \pkg{FiniteDifferences.jl} \citep{FiniteDifferences2021}. 

To offer access to a wider audience, we have packaged the code in the
\proglang{Julia} language as the library
\pkg{BesselK.jl}\footnote{\url{https://github.com/cgeoga/BesselK.jl}}. The
source repository additionally contains example \pkg{R} language bindings as
well as the scripts used to generate the results and figures in this work.

\subsection{Pointwise accuracy} \label{sec:pointwise}
While \pkg{BesselK.jl} is not designed to be a competitor of existing special
functions libraries like \pkg{AMOS} with respect to the accuracy of direct
evaluations of $\besK$, we show that our implementation exhibits competitive
\revision{absolute} accuracy on the entire \revision{testing} domain, both for
direct evaluations as well as derivatives. \revision{Our chosen testing region
is (a dense grid on) $(\nu, x) \in [0.25,10] \times [0.005,30]$, which we pick
for several reasons. For one, the $\nu$ range generously over-covers what we
consider to be the range of sensible values of $\nu$ for fitting Mat\'ern
covariance parameters. With regard to the $x$ domain, we pick $30$ as an upper
end point to verify accuracy in tail values that are small but numerically
relevant.  Moreover, beyond that point any software implementation for $\besK$
will be using an asymptotic expansion, and so naturally there will be good
agreement.} Secondly, we illustrate in log$_{10}$-scale the relative gain of
\pkg{BesselK.jl} over finite difference methods with \pkg{AMOS} for derivative
computations. 

\begin{figure}[!ht]
  \centering
  \begin{tabular}{cc}
\begingroup
  \makeatletter
  \providecommand\color[2][]{%
    \GenericError{(gnuplot) \space\space\space\@spaces}{%
      Package color not loaded in conjunction with
      terminal option `colourtext'%
    }{See the gnuplot documentation for explanation.%
    }{Either use 'blacktext' in gnuplot or load the package
      color.sty in LaTeX.}%
    \renewcommand\color[2][]{}%
  }%
  \providecommand\includegraphics[2][]{%
    \GenericError{(gnuplot) \space\space\space\@spaces}{%
      Package graphicx or graphics not loaded%
    }{See the gnuplot documentation for explanation.%
    }{The gnuplot epslatex terminal needs graphicx.sty or graphics.sty.}%
    \renewcommand\includegraphics[2][]{}%
  }%
  \providecommand\rotatebox[2]{#2}%
  \@ifundefined{ifGPcolor}{%
    \newif\ifGPcolor
    \GPcolortrue
  }{}%
  \@ifundefined{ifGPblacktext}{%
    \newif\ifGPblacktext
    \GPblacktexttrue
  }{}%
  \let\gplgaddtomacro\g@addto@macro
  \gdef\gplbacktext{}%
  \gdef\gplfronttext{}%
  \makeatother
  \ifGPblacktext
    \def\colorrgb#1{}%
    \def\colorgray#1{}%
  \else
    \ifGPcolor
      \def\colorrgb#1{\color[rgb]{#1}}%
      \def\colorgray#1{\color[gray]{#1}}%
      \expandafter\def\csname LTw\endcsname{\color{white}}%
      \expandafter\def\csname LTb\endcsname{\color{black}}%
      \expandafter\def\csname LTa\endcsname{\color{black}}%
      \expandafter\def\csname LT0\endcsname{\color[rgb]{1,0,0}}%
      \expandafter\def\csname LT1\endcsname{\color[rgb]{0,1,0}}%
      \expandafter\def\csname LT2\endcsname{\color[rgb]{0,0,1}}%
      \expandafter\def\csname LT3\endcsname{\color[rgb]{1,0,1}}%
      \expandafter\def\csname LT4\endcsname{\color[rgb]{0,1,1}}%
      \expandafter\def\csname LT5\endcsname{\color[rgb]{1,1,0}}%
      \expandafter\def\csname LT6\endcsname{\color[rgb]{0,0,0}}%
      \expandafter\def\csname LT7\endcsname{\color[rgb]{1,0.3,0}}%
      \expandafter\def\csname LT8\endcsname{\color[rgb]{0.5,0.5,0.5}}%
    \else
      \def\colorrgb#1{\color{black}}%
      \def\colorgray#1{\color[gray]{#1}}%
      \expandafter\def\csname LTw\endcsname{\color{white}}%
      \expandafter\def\csname LTb\endcsname{\color{black}}%
      \expandafter\def\csname LTa\endcsname{\color{black}}%
      \expandafter\def\csname LT0\endcsname{\color{black}}%
      \expandafter\def\csname LT1\endcsname{\color{black}}%
      \expandafter\def\csname LT2\endcsname{\color{black}}%
      \expandafter\def\csname LT3\endcsname{\color{black}}%
      \expandafter\def\csname LT4\endcsname{\color{black}}%
      \expandafter\def\csname LT5\endcsname{\color{black}}%
      \expandafter\def\csname LT6\endcsname{\color{black}}%
      \expandafter\def\csname LT7\endcsname{\color{black}}%
      \expandafter\def\csname LT8\endcsname{\color{black}}%
    \fi
  \fi
    \setlength{\unitlength}{0.0500bp}%
    \ifx\gptboxheight\undefined%
      \newlength{\gptboxheight}%
      \newlength{\gptboxwidth}%
      \newsavebox{\gptboxtext}%
    \fi%
    \setlength{\fboxrule}{0.5pt}%
    \setlength{\fboxsep}{1pt}%
    \definecolor{tbcol}{rgb}{1,1,1}%
    \scalebox{0.6}{
\begin{picture}(7200.00,5040.00)%
    \gplgaddtomacro\gplbacktext{%
      \csname LTb\endcsname
      \put(682,1038){\makebox(0,0)[r]{\strut{}$1$}}%
      \put(682,1456){\makebox(0,0)[r]{\strut{}$2$}}%
      \put(682,1874){\makebox(0,0)[r]{\strut{}$3$}}%
      \put(682,2291){\makebox(0,0)[r]{\strut{}$4$}}%
      \put(682,2709){\makebox(0,0)[r]{\strut{}$5$}}%
      \put(682,3127){\makebox(0,0)[r]{\strut{}$6$}}%
      \put(682,3545){\makebox(0,0)[r]{\strut{}$7$}}%
      \put(682,3963){\makebox(0,0)[r]{\strut{}$8$}}%
      \put(682,4381){\makebox(0,0)[r]{\strut{}$9$}}%
      \put(682,4799){\makebox(0,0)[r]{\strut{}$10$}}%
      \put(1632,484){\makebox(0,0){\strut{}$5$}}%
      \put(2462,484){\makebox(0,0){\strut{}$10$}}%
      \put(3292,484){\makebox(0,0){\strut{}$15$}}%
      \put(4122,484){\makebox(0,0){\strut{}$20$}}%
      \put(4952,484){\makebox(0,0){\strut{}$25$}}%
      \put(5782,484){\makebox(0,0){\strut{}$30$}}%
    }%
    \gplgaddtomacro\gplfronttext{%
      \csname LTb\endcsname
      \put(209,2761){\rotatebox{-270}{\makebox(0,0){\strut{}$\nu$}}}%
      \put(3304,154){\makebox(0,0){\strut{}$x$}}%
      \csname LTb\endcsname
      \put(6300,704){\makebox(0,0)[l]{\strut{}$1\times10^{-20}$}}%
      \put(6300,1137){\makebox(0,0)[l]{\strut{}$1\times10^{-18}$}}%
      \put(6300,1570){\makebox(0,0)[l]{\strut{}$1\times10^{-16}$}}%
      \put(6300,2003){\makebox(0,0)[l]{\strut{}$1\times10^{-14}$}}%
      \put(6300,2436){\makebox(0,0)[l]{\strut{}$1\times10^{-12}$}}%
      \put(6300,2869){\makebox(0,0)[l]{\strut{}$1\times10^{-10}$}}%
      \put(6300,3302){\makebox(0,0)[l]{\strut{}$1\times10^{-8}$}}%
      \put(6300,3736){\makebox(0,0)[l]{\strut{}$1\times10^{-6}$}}%
      \put(6300,4169){\makebox(0,0)[l]{\strut{}$0.0001$}}%
      \put(6300,4602){\makebox(0,0)[l]{\strut{}$0.01$}}%
    }%
    \gplbacktext
    \put(0,0){\includegraphics[width={360.00bp},height={252.00bp}]{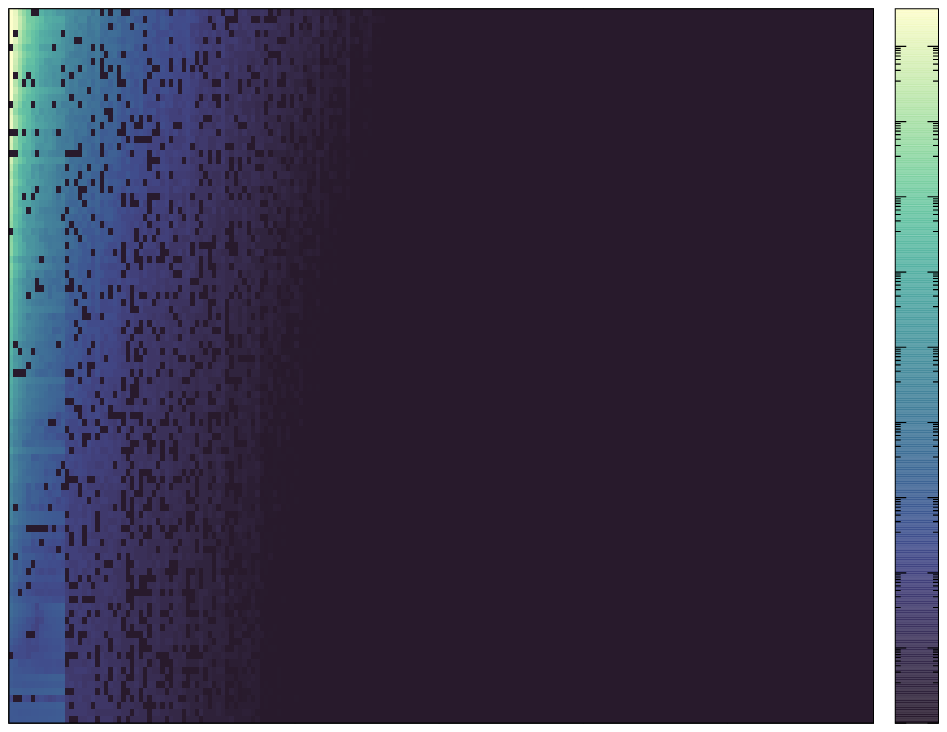}}%
    \gplfronttext
  \end{picture}%
  }
\endgroup
    &
\begingroup
  \makeatletter
  \providecommand\color[2][]{%
    \GenericError{(gnuplot) \space\space\space\@spaces}{%
      Package color not loaded in conjunction with
      terminal option `colourtext'%
    }{See the gnuplot documentation for explanation.%
    }{Either use 'blacktext' in gnuplot or load the package
      color.sty in LaTeX.}%
    \renewcommand\color[2][]{}%
  }%
  \providecommand\includegraphics[2][]{%
    \GenericError{(gnuplot) \space\space\space\@spaces}{%
      Package graphicx or graphics not loaded%
    }{See the gnuplot documentation for explanation.%
    }{The gnuplot epslatex terminal needs graphicx.sty or graphics.sty.}%
    \renewcommand\includegraphics[2][]{}%
  }%
  \providecommand\rotatebox[2]{#2}%
  \@ifundefined{ifGPcolor}{%
    \newif\ifGPcolor
    \GPcolortrue
  }{}%
  \@ifundefined{ifGPblacktext}{%
    \newif\ifGPblacktext
    \GPblacktexttrue
  }{}%
  \let\gplgaddtomacro\g@addto@macro
  \gdef\gplbacktext{}%
  \gdef\gplfronttext{}%
  \makeatother
  \ifGPblacktext
    \def\colorrgb#1{}%
    \def\colorgray#1{}%
  \else
    \ifGPcolor
      \def\colorrgb#1{\color[rgb]{#1}}%
      \def\colorgray#1{\color[gray]{#1}}%
      \expandafter\def\csname LTw\endcsname{\color{white}}%
      \expandafter\def\csname LTb\endcsname{\color{black}}%
      \expandafter\def\csname LTa\endcsname{\color{black}}%
      \expandafter\def\csname LT0\endcsname{\color[rgb]{1,0,0}}%
      \expandafter\def\csname LT1\endcsname{\color[rgb]{0,1,0}}%
      \expandafter\def\csname LT2\endcsname{\color[rgb]{0,0,1}}%
      \expandafter\def\csname LT3\endcsname{\color[rgb]{1,0,1}}%
      \expandafter\def\csname LT4\endcsname{\color[rgb]{0,1,1}}%
      \expandafter\def\csname LT5\endcsname{\color[rgb]{1,1,0}}%
      \expandafter\def\csname LT6\endcsname{\color[rgb]{0,0,0}}%
      \expandafter\def\csname LT7\endcsname{\color[rgb]{1,0.3,0}}%
      \expandafter\def\csname LT8\endcsname{\color[rgb]{0.5,0.5,0.5}}%
    \else
      \def\colorrgb#1{\color{black}}%
      \def\colorgray#1{\color[gray]{#1}}%
      \expandafter\def\csname LTw\endcsname{\color{white}}%
      \expandafter\def\csname LTb\endcsname{\color{black}}%
      \expandafter\def\csname LTa\endcsname{\color{black}}%
      \expandafter\def\csname LT0\endcsname{\color{black}}%
      \expandafter\def\csname LT1\endcsname{\color{black}}%
      \expandafter\def\csname LT2\endcsname{\color{black}}%
      \expandafter\def\csname LT3\endcsname{\color{black}}%
      \expandafter\def\csname LT4\endcsname{\color{black}}%
      \expandafter\def\csname LT5\endcsname{\color{black}}%
      \expandafter\def\csname LT6\endcsname{\color{black}}%
      \expandafter\def\csname LT7\endcsname{\color{black}}%
      \expandafter\def\csname LT8\endcsname{\color{black}}%
    \fi
  \fi
    \setlength{\unitlength}{0.0500bp}%
    \ifx\gptboxheight\undefined%
      \newlength{\gptboxheight}%
      \newlength{\gptboxwidth}%
      \newsavebox{\gptboxtext}%
    \fi%
    \setlength{\fboxrule}{0.5pt}%
    \setlength{\fboxsep}{1pt}%
    \definecolor{tbcol}{rgb}{1,1,1}%
    \scalebox{0.6}{
\begin{picture}(7200.00,5040.00)%
    \gplgaddtomacro\gplbacktext{%
      \csname LTb\endcsname
      \put(682,1038){\makebox(0,0)[r]{\strut{}$1$}}%
      \put(682,1456){\makebox(0,0)[r]{\strut{}$2$}}%
      \put(682,1874){\makebox(0,0)[r]{\strut{}$3$}}%
      \put(682,2291){\makebox(0,0)[r]{\strut{}$4$}}%
      \put(682,2709){\makebox(0,0)[r]{\strut{}$5$}}%
      \put(682,3127){\makebox(0,0)[r]{\strut{}$6$}}%
      \put(682,3545){\makebox(0,0)[r]{\strut{}$7$}}%
      \put(682,3963){\makebox(0,0)[r]{\strut{}$8$}}%
      \put(682,4381){\makebox(0,0)[r]{\strut{}$9$}}%
      \put(682,4799){\makebox(0,0)[r]{\strut{}$10$}}%
      \put(1632,484){\makebox(0,0){\strut{}$5$}}%
      \put(2462,484){\makebox(0,0){\strut{}$10$}}%
      \put(3292,484){\makebox(0,0){\strut{}$15$}}%
      \put(4122,484){\makebox(0,0){\strut{}$20$}}%
      \put(4952,484){\makebox(0,0){\strut{}$25$}}%
      \put(5782,484){\makebox(0,0){\strut{}$30$}}%
    }%
    \gplgaddtomacro\gplfronttext{%
      \csname LTb\endcsname
      \put(209,2761){\rotatebox{-270}{\makebox(0,0){\strut{}$\nu$}}}%
      \put(3304,154){\makebox(0,0){\strut{}$x$}}%
      \csname LTb\endcsname
      \put(6300,704){\makebox(0,0)[l]{\strut{}$1\times10^{-20}$}}%
      \put(6300,1137){\makebox(0,0)[l]{\strut{}$1\times10^{-18}$}}%
      \put(6300,1570){\makebox(0,0)[l]{\strut{}$1\times10^{-16}$}}%
      \put(6300,2003){\makebox(0,0)[l]{\strut{}$1\times10^{-14}$}}%
      \put(6300,2436){\makebox(0,0)[l]{\strut{}$1\times10^{-12}$}}%
      \put(6300,2869){\makebox(0,0)[l]{\strut{}$1\times10^{-10}$}}%
      \put(6300,3302){\makebox(0,0)[l]{\strut{}$1\times10^{-8}$}}%
      \put(6300,3736){\makebox(0,0)[l]{\strut{}$1\times10^{-6}$}}%
      \put(6300,4169){\makebox(0,0)[l]{\strut{}$0.0001$}}%
      \put(6300,4602){\makebox(0,0)[l]{\strut{}$0.01$}}%
    }%
    \gplbacktext
    \put(0,0){\includegraphics[width={360.00bp},height={252.00bp}]{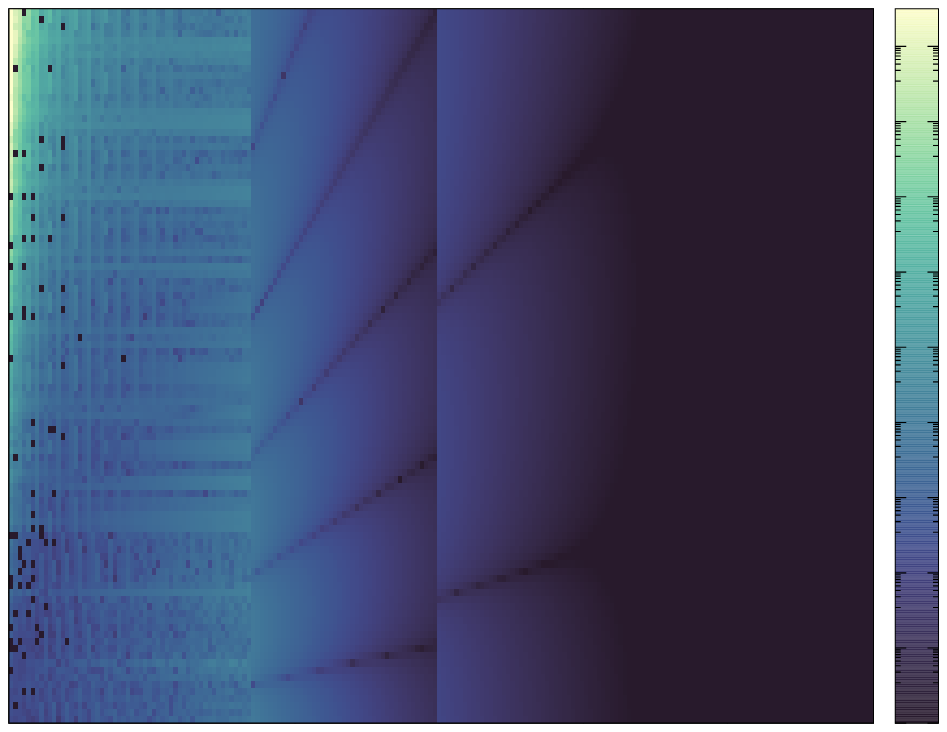}}%
    \gplfronttext
  \end{picture}%
  }
\endgroup
  \end{tabular}
  \caption{Absolute accuracy of direct evaluations with respect to a reference
  solution: \pkg{AMOS} library $\abs{\besK^A - \besK^R}$ (left) and 
  \pkg{BesselK.jl} $\abs{\tilde{\besK} - \besK^R}$ (right).}
  \label{fig:atols}
\end{figure}

\paragraph{Direct evaluations} 
Figure \ref{fig:atols} asseses direct evaluations error for both \pkg{AMOS} and
\pkg{BesselK.jl} using the absolute differences $\abs{\besK^A - \besK^R}$ (left)
and $\abs{\tilde{\besK} - \besK^R}$ (right).  The error behavior is
qualitatively similar for both implementations, with the intermediate argument
range accurate to computer precision, and a small ridge for very small arguments
$x$ and large orders $\nu$ showing increasing inaccuracy. The implementation in
\pkg{BesselK.jl} was tailored to optimize AD computations for different argument
ranges and these ranges are clearly delimited in Figure \ref{fig:atols}. The
loss of accuracy encountered for small to intermediate arguments, approximately
$10^{-12}$ error, is attributed to the difference in numerical treatment and the
choice of truncation levels tuned for accuracy in derivatives. Such a small
accuracy loss in trailing digits will be shown to have a negligible impact on
properties of Mat\'ern covariance matrices relevant to numerical optimization.

\begin{figure}[!ht]
  \centering
  \begin{tabular}{cc}
\begingroup
  \makeatletter
  \providecommand\color[2][]{%
    \GenericError{(gnuplot) \space\space\space\@spaces}{%
      Package color not loaded in conjunction with
      terminal option `colourtext'%
    }{See the gnuplot documentation for explanation.%
    }{Either use 'blacktext' in gnuplot or load the package
      color.sty in LaTeX.}%
    \renewcommand\color[2][]{}%
  }%
  \providecommand\includegraphics[2][]{%
    \GenericError{(gnuplot) \space\space\space\@spaces}{%
      Package graphicx or graphics not loaded%
    }{See the gnuplot documentation for explanation.%
    }{The gnuplot epslatex terminal needs graphicx.sty or graphics.sty.}%
    \renewcommand\includegraphics[2][]{}%
  }%
  \providecommand\rotatebox[2]{#2}%
  \@ifundefined{ifGPcolor}{%
    \newif\ifGPcolor
    \GPcolortrue
  }{}%
  \@ifundefined{ifGPblacktext}{%
    \newif\ifGPblacktext
    \GPblacktexttrue
  }{}%
  \let\gplgaddtomacro\g@addto@macro
  \gdef\gplbacktext{}%
  \gdef\gplfronttext{}%
  \makeatother
  \ifGPblacktext
    \def\colorrgb#1{}%
    \def\colorgray#1{}%
  \else
    \ifGPcolor
      \def\colorrgb#1{\color[rgb]{#1}}%
      \def\colorgray#1{\color[gray]{#1}}%
      \expandafter\def\csname LTw\endcsname{\color{white}}%
      \expandafter\def\csname LTb\endcsname{\color{black}}%
      \expandafter\def\csname LTa\endcsname{\color{black}}%
      \expandafter\def\csname LT0\endcsname{\color[rgb]{1,0,0}}%
      \expandafter\def\csname LT1\endcsname{\color[rgb]{0,1,0}}%
      \expandafter\def\csname LT2\endcsname{\color[rgb]{0,0,1}}%
      \expandafter\def\csname LT3\endcsname{\color[rgb]{1,0,1}}%
      \expandafter\def\csname LT4\endcsname{\color[rgb]{0,1,1}}%
      \expandafter\def\csname LT5\endcsname{\color[rgb]{1,1,0}}%
      \expandafter\def\csname LT6\endcsname{\color[rgb]{0,0,0}}%
      \expandafter\def\csname LT7\endcsname{\color[rgb]{1,0.3,0}}%
      \expandafter\def\csname LT8\endcsname{\color[rgb]{0.5,0.5,0.5}}%
    \else
      \def\colorrgb#1{\color{black}}%
      \def\colorgray#1{\color[gray]{#1}}%
      \expandafter\def\csname LTw\endcsname{\color{white}}%
      \expandafter\def\csname LTb\endcsname{\color{black}}%
      \expandafter\def\csname LTa\endcsname{\color{black}}%
      \expandafter\def\csname LT0\endcsname{\color{black}}%
      \expandafter\def\csname LT1\endcsname{\color{black}}%
      \expandafter\def\csname LT2\endcsname{\color{black}}%
      \expandafter\def\csname LT3\endcsname{\color{black}}%
      \expandafter\def\csname LT4\endcsname{\color{black}}%
      \expandafter\def\csname LT5\endcsname{\color{black}}%
      \expandafter\def\csname LT6\endcsname{\color{black}}%
      \expandafter\def\csname LT7\endcsname{\color{black}}%
      \expandafter\def\csname LT8\endcsname{\color{black}}%
    \fi
  \fi
    \setlength{\unitlength}{0.0500bp}%
    \ifx\gptboxheight\undefined%
      \newlength{\gptboxheight}%
      \newlength{\gptboxwidth}%
      \newsavebox{\gptboxtext}%
    \fi%
    \setlength{\fboxrule}{0.5pt}%
    \setlength{\fboxsep}{1pt}%
    \definecolor{tbcol}{rgb}{1,1,1}%
    \scalebox{0.6}{
\begin{picture}(7200.00,5040.00)%
    \gplgaddtomacro\gplbacktext{%
      \csname LTb\endcsname
      \put(682,1038){\makebox(0,0)[r]{\strut{}$1$}}%
      \put(682,1456){\makebox(0,0)[r]{\strut{}$2$}}%
      \put(682,1874){\makebox(0,0)[r]{\strut{}$3$}}%
      \put(682,2291){\makebox(0,0)[r]{\strut{}$4$}}%
      \put(682,2709){\makebox(0,0)[r]{\strut{}$5$}}%
      \put(682,3127){\makebox(0,0)[r]{\strut{}$6$}}%
      \put(682,3545){\makebox(0,0)[r]{\strut{}$7$}}%
      \put(682,3963){\makebox(0,0)[r]{\strut{}$8$}}%
      \put(682,4381){\makebox(0,0)[r]{\strut{}$9$}}%
      \put(682,4799){\makebox(0,0)[r]{\strut{}$10$}}%
      \put(1632,484){\makebox(0,0){\strut{}$5$}}%
      \put(2462,484){\makebox(0,0){\strut{}$10$}}%
      \put(3292,484){\makebox(0,0){\strut{}$15$}}%
      \put(4122,484){\makebox(0,0){\strut{}$20$}}%
      \put(4952,484){\makebox(0,0){\strut{}$25$}}%
      \put(5782,484){\makebox(0,0){\strut{}$30$}}%
    }%
    \gplgaddtomacro\gplfronttext{%
      \csname LTb\endcsname
      \put(209,2761){\rotatebox{-270}{\makebox(0,0){\strut{}$\nu$}}}%
      \put(3304,154){\makebox(0,0){\strut{}$x$}}%
      \csname LTb\endcsname
      \put(6300,704){\makebox(0,0)[l]{\strut{}$1\times10^{-20}$}}%
      \put(6300,1526){\makebox(0,0)[l]{\strut{}$1\times10^{-15}$}}%
      \put(6300,2350){\makebox(0,0)[l]{\strut{}$1\times10^{-10}$}}%
      \put(6300,3173){\makebox(0,0)[l]{\strut{}$1\times10^{-5}$}}%
      \put(6300,3996){\makebox(0,0)[l]{\strut{}$1$}}%
      \put(6300,4819){\makebox(0,0)[l]{\strut{}$100000$}}%
    }%
    \gplbacktext
    \put(0,0){\includegraphics[width={360.00bp},height={252.00bp}]{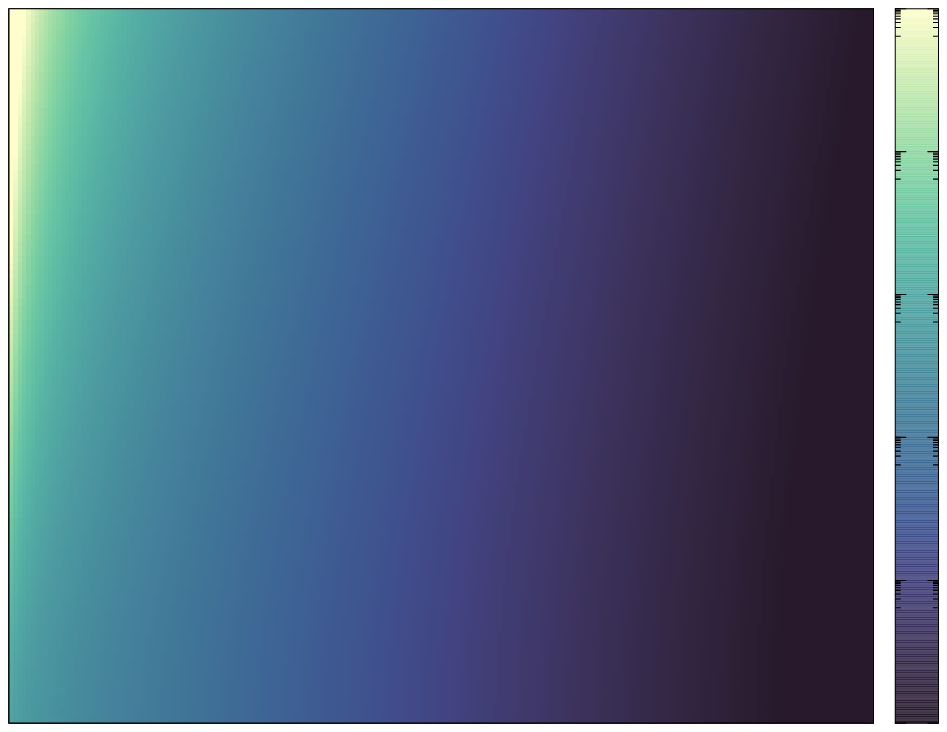}}%
    \gplfronttext
  \end{picture}%
  }
\endgroup
    &
\begingroup
  \makeatletter
  \providecommand\color[2][]{%
    \GenericError{(gnuplot) \space\space\space\@spaces}{%
      Package color not loaded in conjunction with
      terminal option `colourtext'%
    }{See the gnuplot documentation for explanation.%
    }{Either use 'blacktext' in gnuplot or load the package
      color.sty in LaTeX.}%
    \renewcommand\color[2][]{}%
  }%
  \providecommand\includegraphics[2][]{%
    \GenericError{(gnuplot) \space\space\space\@spaces}{%
      Package graphicx or graphics not loaded%
    }{See the gnuplot documentation for explanation.%
    }{The gnuplot epslatex terminal needs graphicx.sty or graphics.sty.}%
    \renewcommand\includegraphics[2][]{}%
  }%
  \providecommand\rotatebox[2]{#2}%
  \@ifundefined{ifGPcolor}{%
    \newif\ifGPcolor
    \GPcolortrue
  }{}%
  \@ifundefined{ifGPblacktext}{%
    \newif\ifGPblacktext
    \GPblacktexttrue
  }{}%
  \let\gplgaddtomacro\g@addto@macro
  \gdef\gplbacktext{}%
  \gdef\gplfronttext{}%
  \makeatother
  \ifGPblacktext
    \def\colorrgb#1{}%
    \def\colorgray#1{}%
  \else
    \ifGPcolor
      \def\colorrgb#1{\color[rgb]{#1}}%
      \def\colorgray#1{\color[gray]{#1}}%
      \expandafter\def\csname LTw\endcsname{\color{white}}%
      \expandafter\def\csname LTb\endcsname{\color{black}}%
      \expandafter\def\csname LTa\endcsname{\color{black}}%
      \expandafter\def\csname LT0\endcsname{\color[rgb]{1,0,0}}%
      \expandafter\def\csname LT1\endcsname{\color[rgb]{0,1,0}}%
      \expandafter\def\csname LT2\endcsname{\color[rgb]{0,0,1}}%
      \expandafter\def\csname LT3\endcsname{\color[rgb]{1,0,1}}%
      \expandafter\def\csname LT4\endcsname{\color[rgb]{0,1,1}}%
      \expandafter\def\csname LT5\endcsname{\color[rgb]{1,1,0}}%
      \expandafter\def\csname LT6\endcsname{\color[rgb]{0,0,0}}%
      \expandafter\def\csname LT7\endcsname{\color[rgb]{1,0.3,0}}%
      \expandafter\def\csname LT8\endcsname{\color[rgb]{0.5,0.5,0.5}}%
    \else
      \def\colorrgb#1{\color{black}}%
      \def\colorgray#1{\color[gray]{#1}}%
      \expandafter\def\csname LTw\endcsname{\color{white}}%
      \expandafter\def\csname LTb\endcsname{\color{black}}%
      \expandafter\def\csname LTa\endcsname{\color{black}}%
      \expandafter\def\csname LT0\endcsname{\color{black}}%
      \expandafter\def\csname LT1\endcsname{\color{black}}%
      \expandafter\def\csname LT2\endcsname{\color{black}}%
      \expandafter\def\csname LT3\endcsname{\color{black}}%
      \expandafter\def\csname LT4\endcsname{\color{black}}%
      \expandafter\def\csname LT5\endcsname{\color{black}}%
      \expandafter\def\csname LT6\endcsname{\color{black}}%
      \expandafter\def\csname LT7\endcsname{\color{black}}%
      \expandafter\def\csname LT8\endcsname{\color{black}}%
    \fi
  \fi
    \setlength{\unitlength}{0.0500bp}%
    \ifx\gptboxheight\undefined%
      \newlength{\gptboxheight}%
      \newlength{\gptboxwidth}%
      \newsavebox{\gptboxtext}%
    \fi%
    \setlength{\fboxrule}{0.5pt}%
    \setlength{\fboxsep}{1pt}%
    \definecolor{tbcol}{rgb}{1,1,1}%
    \scalebox{0.6}{
\begin{picture}(7200.00,5040.00)%
    \gplgaddtomacro\gplbacktext{%
      \csname LTb\endcsname
      \put(682,1038){\makebox(0,0)[r]{\strut{}$1$}}%
      \put(682,1456){\makebox(0,0)[r]{\strut{}$2$}}%
      \put(682,1874){\makebox(0,0)[r]{\strut{}$3$}}%
      \put(682,2291){\makebox(0,0)[r]{\strut{}$4$}}%
      \put(682,2709){\makebox(0,0)[r]{\strut{}$5$}}%
      \put(682,3127){\makebox(0,0)[r]{\strut{}$6$}}%
      \put(682,3545){\makebox(0,0)[r]{\strut{}$7$}}%
      \put(682,3963){\makebox(0,0)[r]{\strut{}$8$}}%
      \put(682,4381){\makebox(0,0)[r]{\strut{}$9$}}%
      \put(682,4799){\makebox(0,0)[r]{\strut{}$10$}}%
      \put(1632,484){\makebox(0,0){\strut{}$5$}}%
      \put(2462,484){\makebox(0,0){\strut{}$10$}}%
      \put(3292,484){\makebox(0,0){\strut{}$15$}}%
      \put(4122,484){\makebox(0,0){\strut{}$20$}}%
      \put(4952,484){\makebox(0,0){\strut{}$25$}}%
      \put(5782,484){\makebox(0,0){\strut{}$30$}}%
    }%
    \gplgaddtomacro\gplfronttext{%
      \csname LTb\endcsname
      \put(209,2761){\rotatebox{-270}{\makebox(0,0){\strut{}$\nu$}}}%
      \put(3304,154){\makebox(0,0){\strut{}$x$}}%
      \csname LTb\endcsname
      \put(6300,704){\makebox(0,0)[l]{\strut{}$1\times10^{-20}$}}%
      \put(6300,1526){\makebox(0,0)[l]{\strut{}$1\times10^{-15}$}}%
      \put(6300,2350){\makebox(0,0)[l]{\strut{}$1\times10^{-10}$}}%
      \put(6300,3173){\makebox(0,0)[l]{\strut{}$1\times10^{-5}$}}%
      \put(6300,3996){\makebox(0,0)[l]{\strut{}$1$}}%
      \put(6300,4819){\makebox(0,0)[l]{\strut{}$100000$}}%
    }%
    \gplbacktext
    \put(0,0){\includegraphics[width={360.00bp},height={252.00bp}]{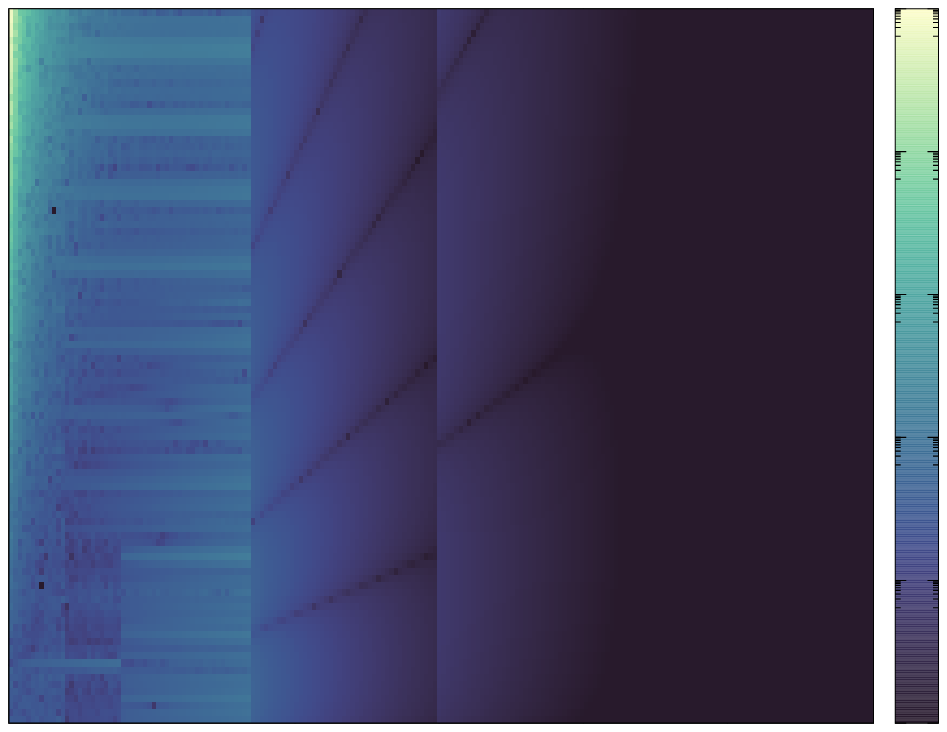}}%
    \gplfronttext
  \end{picture}%
  }
\endgroup
  \end{tabular}
  \caption{Absolute accuracy of first derivatives with respect to a reference solution: \pkg{AMOS} using finite differences 
  $\abs{\partial_\nu^{1\text{FD}} \besK - \partial_\nu^{1\text{R}} \besK}$ (left)
and \pkg{BesselK.jl} using AD
  $\abs{\partial_\nu^{1\text{AD}} \tilde{\besK}  - \partial_\nu^{1\text{R}} \besK}$
  (right).}
  \label{fig:atols_deriv}
\end{figure}

\begin{figure}[!h]
  \centering
  \begin{tabular}{cc}
\begingroup
  \makeatletter
  \providecommand\color[2][]{%
    \GenericError{(gnuplot) \space\space\space\@spaces}{%
      Package color not loaded in conjunction with
      terminal option `colourtext'%
    }{See the gnuplot documentation for explanation.%
    }{Either use 'blacktext' in gnuplot or load the package
      color.sty in LaTeX.}%
    \renewcommand\color[2][]{}%
  }%
  \providecommand\includegraphics[2][]{%
    \GenericError{(gnuplot) \space\space\space\@spaces}{%
      Package graphicx or graphics not loaded%
    }{See the gnuplot documentation for explanation.%
    }{The gnuplot epslatex terminal needs graphicx.sty or graphics.sty.}%
    \renewcommand\includegraphics[2][]{}%
  }%
  \providecommand\rotatebox[2]{#2}%
  \@ifundefined{ifGPcolor}{%
    \newif\ifGPcolor
    \GPcolortrue
  }{}%
  \@ifundefined{ifGPblacktext}{%
    \newif\ifGPblacktext
    \GPblacktexttrue
  }{}%
  \let\gplgaddtomacro\g@addto@macro
  \gdef\gplbacktext{}%
  \gdef\gplfronttext{}%
  \makeatother
  \ifGPblacktext
    \def\colorrgb#1{}%
    \def\colorgray#1{}%
  \else
    \ifGPcolor
      \def\colorrgb#1{\color[rgb]{#1}}%
      \def\colorgray#1{\color[gray]{#1}}%
      \expandafter\def\csname LTw\endcsname{\color{white}}%
      \expandafter\def\csname LTb\endcsname{\color{black}}%
      \expandafter\def\csname LTa\endcsname{\color{black}}%
      \expandafter\def\csname LT0\endcsname{\color[rgb]{1,0,0}}%
      \expandafter\def\csname LT1\endcsname{\color[rgb]{0,1,0}}%
      \expandafter\def\csname LT2\endcsname{\color[rgb]{0,0,1}}%
      \expandafter\def\csname LT3\endcsname{\color[rgb]{1,0,1}}%
      \expandafter\def\csname LT4\endcsname{\color[rgb]{0,1,1}}%
      \expandafter\def\csname LT5\endcsname{\color[rgb]{1,1,0}}%
      \expandafter\def\csname LT6\endcsname{\color[rgb]{0,0,0}}%
      \expandafter\def\csname LT7\endcsname{\color[rgb]{1,0.3,0}}%
      \expandafter\def\csname LT8\endcsname{\color[rgb]{0.5,0.5,0.5}}%
    \else
      \def\colorrgb#1{\color{black}}%
      \def\colorgray#1{\color[gray]{#1}}%
      \expandafter\def\csname LTw\endcsname{\color{white}}%
      \expandafter\def\csname LTb\endcsname{\color{black}}%
      \expandafter\def\csname LTa\endcsname{\color{black}}%
      \expandafter\def\csname LT0\endcsname{\color{black}}%
      \expandafter\def\csname LT1\endcsname{\color{black}}%
      \expandafter\def\csname LT2\endcsname{\color{black}}%
      \expandafter\def\csname LT3\endcsname{\color{black}}%
      \expandafter\def\csname LT4\endcsname{\color{black}}%
      \expandafter\def\csname LT5\endcsname{\color{black}}%
      \expandafter\def\csname LT6\endcsname{\color{black}}%
      \expandafter\def\csname LT7\endcsname{\color{black}}%
      \expandafter\def\csname LT8\endcsname{\color{black}}%
    \fi
  \fi
    \setlength{\unitlength}{0.0500bp}%
    \ifx\gptboxheight\undefined%
      \newlength{\gptboxheight}%
      \newlength{\gptboxwidth}%
      \newsavebox{\gptboxtext}%
    \fi%
    \setlength{\fboxrule}{0.5pt}%
    \setlength{\fboxsep}{1pt}%
    \definecolor{tbcol}{rgb}{1,1,1}%
    \scalebox{0.6}{
\begin{picture}(7200.00,5040.00)%
    \gplgaddtomacro\gplbacktext{%
      \csname LTb\endcsname
      \put(682,1038){\makebox(0,0)[r]{\strut{}$1$}}%
      \put(682,1456){\makebox(0,0)[r]{\strut{}$2$}}%
      \put(682,1874){\makebox(0,0)[r]{\strut{}$3$}}%
      \put(682,2291){\makebox(0,0)[r]{\strut{}$4$}}%
      \put(682,2709){\makebox(0,0)[r]{\strut{}$5$}}%
      \put(682,3127){\makebox(0,0)[r]{\strut{}$6$}}%
      \put(682,3545){\makebox(0,0)[r]{\strut{}$7$}}%
      \put(682,3963){\makebox(0,0)[r]{\strut{}$8$}}%
      \put(682,4381){\makebox(0,0)[r]{\strut{}$9$}}%
      \put(682,4799){\makebox(0,0)[r]{\strut{}$10$}}%
      \put(1632,484){\makebox(0,0){\strut{}$5$}}%
      \put(2462,484){\makebox(0,0){\strut{}$10$}}%
      \put(3292,484){\makebox(0,0){\strut{}$15$}}%
      \put(4122,484){\makebox(0,0){\strut{}$20$}}%
      \put(4952,484){\makebox(0,0){\strut{}$25$}}%
      \put(5782,484){\makebox(0,0){\strut{}$30$}}%
    }%
    \gplgaddtomacro\gplfronttext{%
      \csname LTb\endcsname
      \put(209,2761){\rotatebox{-270}{\makebox(0,0){\strut{}$\nu$}}}%
      \put(3304,154){\makebox(0,0){\strut{}$x$}}%
      \csname LTb\endcsname
      \put(6300,704){\makebox(0,0)[l]{\strut{}$1\times10^{-20}$}}%
      \put(6300,1340){\makebox(0,0)[l]{\strut{}$1\times10^{-15}$}}%
      \put(6300,1977){\makebox(0,0)[l]{\strut{}$1\times10^{-10}$}}%
      \put(6300,2614){\makebox(0,0)[l]{\strut{}$1\times10^{-5}$}}%
      \put(6300,3251){\makebox(0,0)[l]{\strut{}$1$}}%
      \put(6300,3888){\makebox(0,0)[l]{\strut{}$100000$}}%
      \put(6300,4525){\makebox(0,0)[l]{\strut{}$1\times10^{10}$}}%
    }%
    \gplbacktext
    \put(0,0){\includegraphics[width={360.00bp},height={252.00bp}]{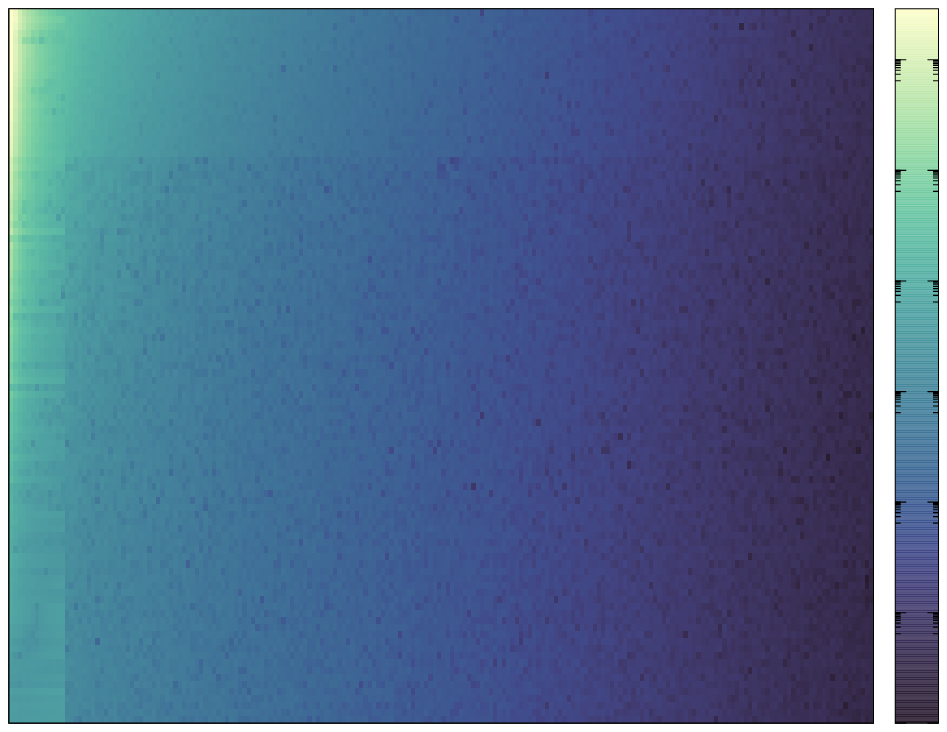}}%
    \gplfronttext
  \end{picture}%
  }
\endgroup
    &
\begingroup
  \makeatletter
  \providecommand\color[2][]{%
    \GenericError{(gnuplot) \space\space\space\@spaces}{%
      Package color not loaded in conjunction with
      terminal option `colourtext'%
    }{See the gnuplot documentation for explanation.%
    }{Either use 'blacktext' in gnuplot or load the package
      color.sty in LaTeX.}%
    \renewcommand\color[2][]{}%
  }%
  \providecommand\includegraphics[2][]{%
    \GenericError{(gnuplot) \space\space\space\@spaces}{%
      Package graphicx or graphics not loaded%
    }{See the gnuplot documentation for explanation.%
    }{The gnuplot epslatex terminal needs graphicx.sty or graphics.sty.}%
    \renewcommand\includegraphics[2][]{}%
  }%
  \providecommand\rotatebox[2]{#2}%
  \@ifundefined{ifGPcolor}{%
    \newif\ifGPcolor
    \GPcolortrue
  }{}%
  \@ifundefined{ifGPblacktext}{%
    \newif\ifGPblacktext
    \GPblacktexttrue
  }{}%
  \let\gplgaddtomacro\g@addto@macro
  \gdef\gplbacktext{}%
  \gdef\gplfronttext{}%
  \makeatother
  \ifGPblacktext
    \def\colorrgb#1{}%
    \def\colorgray#1{}%
  \else
    \ifGPcolor
      \def\colorrgb#1{\color[rgb]{#1}}%
      \def\colorgray#1{\color[gray]{#1}}%
      \expandafter\def\csname LTw\endcsname{\color{white}}%
      \expandafter\def\csname LTb\endcsname{\color{black}}%
      \expandafter\def\csname LTa\endcsname{\color{black}}%
      \expandafter\def\csname LT0\endcsname{\color[rgb]{1,0,0}}%
      \expandafter\def\csname LT1\endcsname{\color[rgb]{0,1,0}}%
      \expandafter\def\csname LT2\endcsname{\color[rgb]{0,0,1}}%
      \expandafter\def\csname LT3\endcsname{\color[rgb]{1,0,1}}%
      \expandafter\def\csname LT4\endcsname{\color[rgb]{0,1,1}}%
      \expandafter\def\csname LT5\endcsname{\color[rgb]{1,1,0}}%
      \expandafter\def\csname LT6\endcsname{\color[rgb]{0,0,0}}%
      \expandafter\def\csname LT7\endcsname{\color[rgb]{1,0.3,0}}%
      \expandafter\def\csname LT8\endcsname{\color[rgb]{0.5,0.5,0.5}}%
    \else
      \def\colorrgb#1{\color{black}}%
      \def\colorgray#1{\color[gray]{#1}}%
      \expandafter\def\csname LTw\endcsname{\color{white}}%
      \expandafter\def\csname LTb\endcsname{\color{black}}%
      \expandafter\def\csname LTa\endcsname{\color{black}}%
      \expandafter\def\csname LT0\endcsname{\color{black}}%
      \expandafter\def\csname LT1\endcsname{\color{black}}%
      \expandafter\def\csname LT2\endcsname{\color{black}}%
      \expandafter\def\csname LT3\endcsname{\color{black}}%
      \expandafter\def\csname LT4\endcsname{\color{black}}%
      \expandafter\def\csname LT5\endcsname{\color{black}}%
      \expandafter\def\csname LT6\endcsname{\color{black}}%
      \expandafter\def\csname LT7\endcsname{\color{black}}%
      \expandafter\def\csname LT8\endcsname{\color{black}}%
    \fi
  \fi
    \setlength{\unitlength}{0.0500bp}%
    \ifx\gptboxheight\undefined%
      \newlength{\gptboxheight}%
      \newlength{\gptboxwidth}%
      \newsavebox{\gptboxtext}%
    \fi%
    \setlength{\fboxrule}{0.5pt}%
    \setlength{\fboxsep}{1pt}%
    \definecolor{tbcol}{rgb}{1,1,1}%
    \scalebox{0.6}{
\begin{picture}(7200.00,5040.00)%
    \gplgaddtomacro\gplbacktext{%
      \csname LTb\endcsname
      \put(682,1038){\makebox(0,0)[r]{\strut{}$1$}}%
      \put(682,1456){\makebox(0,0)[r]{\strut{}$2$}}%
      \put(682,1874){\makebox(0,0)[r]{\strut{}$3$}}%
      \put(682,2291){\makebox(0,0)[r]{\strut{}$4$}}%
      \put(682,2709){\makebox(0,0)[r]{\strut{}$5$}}%
      \put(682,3127){\makebox(0,0)[r]{\strut{}$6$}}%
      \put(682,3545){\makebox(0,0)[r]{\strut{}$7$}}%
      \put(682,3963){\makebox(0,0)[r]{\strut{}$8$}}%
      \put(682,4381){\makebox(0,0)[r]{\strut{}$9$}}%
      \put(682,4799){\makebox(0,0)[r]{\strut{}$10$}}%
      \put(1632,484){\makebox(0,0){\strut{}$5$}}%
      \put(2462,484){\makebox(0,0){\strut{}$10$}}%
      \put(3292,484){\makebox(0,0){\strut{}$15$}}%
      \put(4122,484){\makebox(0,0){\strut{}$20$}}%
      \put(4952,484){\makebox(0,0){\strut{}$25$}}%
      \put(5782,484){\makebox(0,0){\strut{}$30$}}%
    }%
    \gplgaddtomacro\gplfronttext{%
      \csname LTb\endcsname
      \put(209,2761){\rotatebox{-270}{\makebox(0,0){\strut{}$\nu$}}}%
      \put(3304,154){\makebox(0,0){\strut{}$x$}}%
      \csname LTb\endcsname
      \put(6300,704){\makebox(0,0)[l]{\strut{}$1\times10^{-20}$}}%
      \put(6300,1340){\makebox(0,0)[l]{\strut{}$1\times10^{-15}$}}%
      \put(6300,1977){\makebox(0,0)[l]{\strut{}$1\times10^{-10}$}}%
      \put(6300,2614){\makebox(0,0)[l]{\strut{}$1\times10^{-5}$}}%
      \put(6300,3251){\makebox(0,0)[l]{\strut{}$1$}}%
      \put(6300,3888){\makebox(0,0)[l]{\strut{}$100000$}}%
      \put(6300,4525){\makebox(0,0)[l]{\strut{}$1\times10^{10}$}}%
    }%
    \gplbacktext
    \put(0,0){\includegraphics[width={360.00bp},height={252.00bp}]{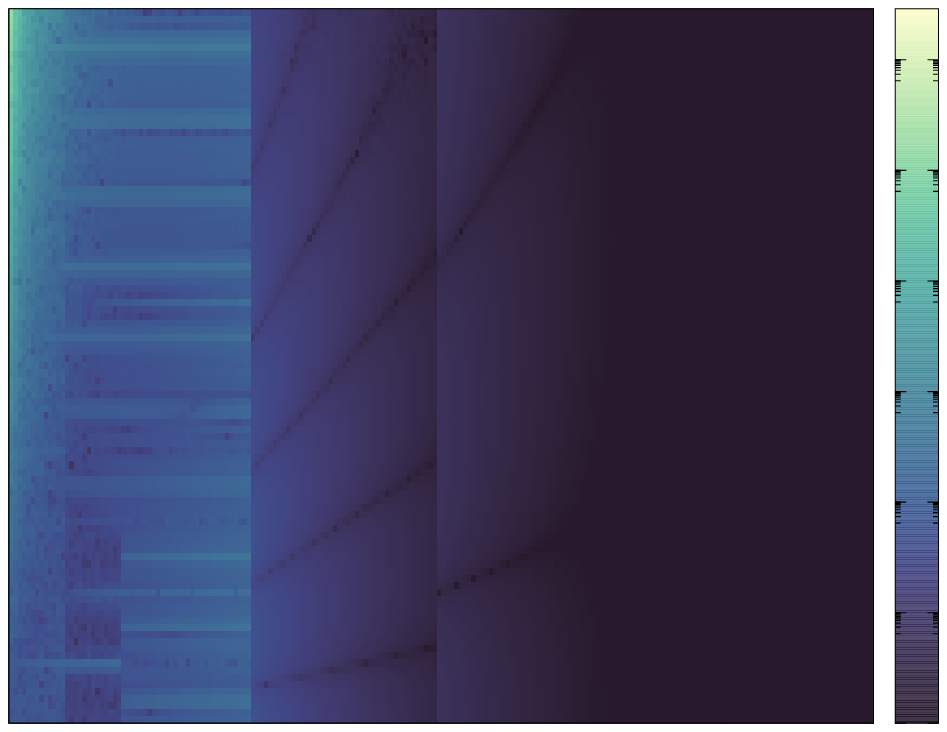}}%
    \gplfronttext
  \end{picture}%
  }
\endgroup
  \end{tabular}
  \caption{Absolute accuracy of second derivatives with respect to a reference solution: \pkg{AMOS} using finite differences 
  $\abs{\partial_\nu^{2\text{FD}} \besK - \partial_\nu^{2\text{R}} \besK}$ (left)
and \pkg{BesselK.jl} using AD
  $\abs{\partial_\nu^{2\text{AD}} \tilde{\besK}  - \partial_\nu^{2\text{R}} \besK}$
  (right).}
  \label{fig:atols_deriv2}
\end{figure}

\paragraph{Derivatives} 
For first derivative computations, illustrated in Figure \ref{fig:atols_deriv},
we see that \pkg{BesselK.jl} (right) is noticeably superior to finite difference
estimates with \pkg{AMOS} (left) in the small argument range. The latter method
incurs rather large errors well within the intermediate argument range $x>8.5$,
while \pkg{BesselK.jl} consistently out-performs as the argument
increases.  As higher-order derivatives are taken, as illustrated in
Figure~\ref{fig:atols_deriv2} for second derivatives, this behavior
becomes even more pronounced and automatic differentiation clearly overtakes
finite difference computations by a significant margin.

\begin{figure}[!ht]
  \centering
  \begin{tabular}{cc}
\begingroup
  \makeatletter
  \providecommand\color[2][]{%
    \GenericError{(gnuplot) \space\space\space\@spaces}{%
      Package color not loaded in conjunction with
      terminal option `colourtext'%
    }{See the gnuplot documentation for explanation.%
    }{Either use 'blacktext' in gnuplot or load the package
      color.sty in LaTeX.}%
    \renewcommand\color[2][]{}%
  }%
  \providecommand\includegraphics[2][]{%
    \GenericError{(gnuplot) \space\space\space\@spaces}{%
      Package graphicx or graphics not loaded%
    }{See the gnuplot documentation for explanation.%
    }{The gnuplot epslatex terminal needs graphicx.sty or graphics.sty.}%
    \renewcommand\includegraphics[2][]{}%
  }%
  \providecommand\rotatebox[2]{#2}%
  \@ifundefined{ifGPcolor}{%
    \newif\ifGPcolor
    \GPcolortrue
  }{}%
  \@ifundefined{ifGPblacktext}{%
    \newif\ifGPblacktext
    \GPblacktexttrue
  }{}%
  \let\gplgaddtomacro\g@addto@macro
  \gdef\gplbacktext{}%
  \gdef\gplfronttext{}%
  \makeatother
  \ifGPblacktext
    \def\colorrgb#1{}%
    \def\colorgray#1{}%
  \else
    \ifGPcolor
      \def\colorrgb#1{\color[rgb]{#1}}%
      \def\colorgray#1{\color[gray]{#1}}%
      \expandafter\def\csname LTw\endcsname{\color{white}}%
      \expandafter\def\csname LTb\endcsname{\color{black}}%
      \expandafter\def\csname LTa\endcsname{\color{black}}%
      \expandafter\def\csname LT0\endcsname{\color[rgb]{1,0,0}}%
      \expandafter\def\csname LT1\endcsname{\color[rgb]{0,1,0}}%
      \expandafter\def\csname LT2\endcsname{\color[rgb]{0,0,1}}%
      \expandafter\def\csname LT3\endcsname{\color[rgb]{1,0,1}}%
      \expandafter\def\csname LT4\endcsname{\color[rgb]{0,1,1}}%
      \expandafter\def\csname LT5\endcsname{\color[rgb]{1,1,0}}%
      \expandafter\def\csname LT6\endcsname{\color[rgb]{0,0,0}}%
      \expandafter\def\csname LT7\endcsname{\color[rgb]{1,0.3,0}}%
      \expandafter\def\csname LT8\endcsname{\color[rgb]{0.5,0.5,0.5}}%
    \else
      \def\colorrgb#1{\color{black}}%
      \def\colorgray#1{\color[gray]{#1}}%
      \expandafter\def\csname LTw\endcsname{\color{white}}%
      \expandafter\def\csname LTb\endcsname{\color{black}}%
      \expandafter\def\csname LTa\endcsname{\color{black}}%
      \expandafter\def\csname LT0\endcsname{\color{black}}%
      \expandafter\def\csname LT1\endcsname{\color{black}}%
      \expandafter\def\csname LT2\endcsname{\color{black}}%
      \expandafter\def\csname LT3\endcsname{\color{black}}%
      \expandafter\def\csname LT4\endcsname{\color{black}}%
      \expandafter\def\csname LT5\endcsname{\color{black}}%
      \expandafter\def\csname LT6\endcsname{\color{black}}%
      \expandafter\def\csname LT7\endcsname{\color{black}}%
      \expandafter\def\csname LT8\endcsname{\color{black}}%
    \fi
  \fi
    \setlength{\unitlength}{0.0500bp}%
    \ifx\gptboxheight\undefined%
      \newlength{\gptboxheight}%
      \newlength{\gptboxwidth}%
      \newsavebox{\gptboxtext}%
    \fi%
    \setlength{\fboxrule}{0.5pt}%
    \setlength{\fboxsep}{1pt}%
    \definecolor{tbcol}{rgb}{1,1,1}%
    \scalebox{0.6}{
\begin{picture}(7200.00,5040.00)%
    \gplgaddtomacro\gplbacktext{%
      \csname LTb\endcsname
      \put(682,1038){\makebox(0,0)[r]{\strut{}$1$}}%
      \put(682,1456){\makebox(0,0)[r]{\strut{}$2$}}%
      \put(682,1874){\makebox(0,0)[r]{\strut{}$3$}}%
      \put(682,2291){\makebox(0,0)[r]{\strut{}$4$}}%
      \put(682,2709){\makebox(0,0)[r]{\strut{}$5$}}%
      \put(682,3127){\makebox(0,0)[r]{\strut{}$6$}}%
      \put(682,3545){\makebox(0,0)[r]{\strut{}$7$}}%
      \put(682,3963){\makebox(0,0)[r]{\strut{}$8$}}%
      \put(682,4381){\makebox(0,0)[r]{\strut{}$9$}}%
      \put(682,4799){\makebox(0,0)[r]{\strut{}$10$}}%
      \put(1632,484){\makebox(0,0){\strut{}$5$}}%
      \put(2462,484){\makebox(0,0){\strut{}$10$}}%
      \put(3292,484){\makebox(0,0){\strut{}$15$}}%
      \put(4122,484){\makebox(0,0){\strut{}$20$}}%
      \put(4952,484){\makebox(0,0){\strut{}$25$}}%
      \put(5782,484){\makebox(0,0){\strut{}$30$}}%
    }%
    \gplgaddtomacro\gplfronttext{%
      \csname LTb\endcsname
      \put(209,2761){\rotatebox{-270}{\makebox(0,0){\strut{}$\nu$}}}%
      \put(3304,154){\makebox(0,0){\strut{}$x$}}%
      \csname LTb\endcsname
      \put(6300,704){\makebox(0,0)[l]{\strut{}$-10$}}%
      \put(6300,1732){\makebox(0,0)[l]{\strut{}$-5$}}%
      \put(6300,2761){\makebox(0,0)[l]{\strut{}$0$}}%
      \put(6300,3790){\makebox(0,0)[l]{\strut{}$5$}}%
      \put(6300,4819){\makebox(0,0)[l]{\strut{}$10$}}%
    }%
    \gplbacktext
    \put(0,0){\includegraphics[width={360.00bp},height={252.00bp}]{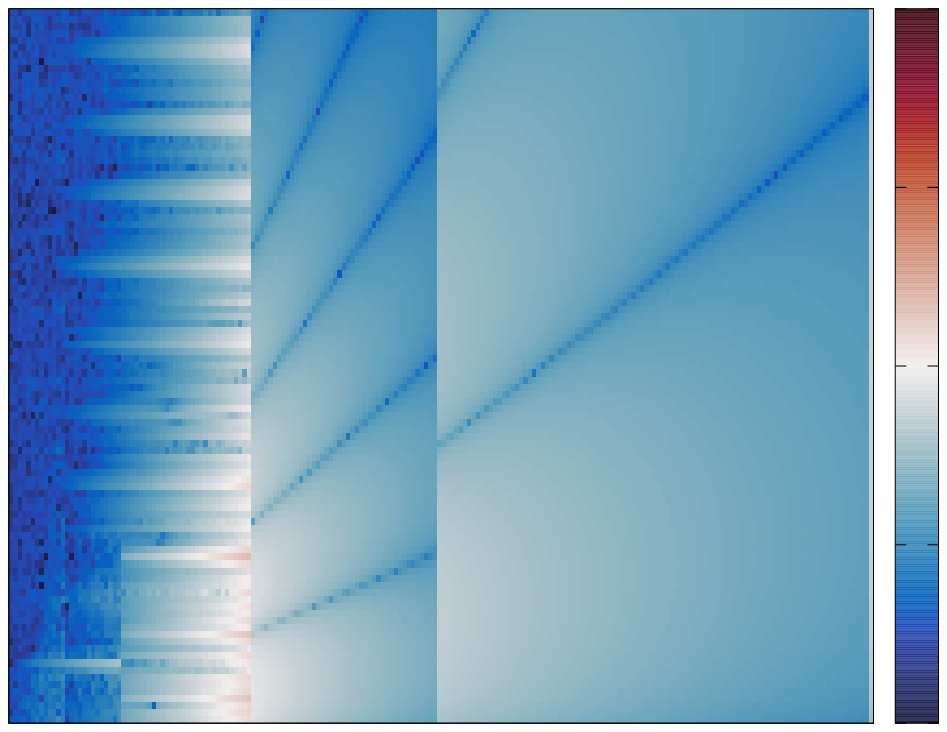}}%
    \gplfronttext
  \end{picture}%
  }
\endgroup
    &
\begingroup
  \makeatletter
  \providecommand\color[2][]{%
    \GenericError{(gnuplot) \space\space\space\@spaces}{%
      Package color not loaded in conjunction with
      terminal option `colourtext'%
    }{See the gnuplot documentation for explanation.%
    }{Either use 'blacktext' in gnuplot or load the package
      color.sty in LaTeX.}%
    \renewcommand\color[2][]{}%
  }%
  \providecommand\includegraphics[2][]{%
    \GenericError{(gnuplot) \space\space\space\@spaces}{%
      Package graphicx or graphics not loaded%
    }{See the gnuplot documentation for explanation.%
    }{The gnuplot epslatex terminal needs graphicx.sty or graphics.sty.}%
    \renewcommand\includegraphics[2][]{}%
  }%
  \providecommand\rotatebox[2]{#2}%
  \@ifundefined{ifGPcolor}{%
    \newif\ifGPcolor
    \GPcolortrue
  }{}%
  \@ifundefined{ifGPblacktext}{%
    \newif\ifGPblacktext
    \GPblacktexttrue
  }{}%
  \let\gplgaddtomacro\g@addto@macro
  \gdef\gplbacktext{}%
  \gdef\gplfronttext{}%
  \makeatother
  \ifGPblacktext
    \def\colorrgb#1{}%
    \def\colorgray#1{}%
  \else
    \ifGPcolor
      \def\colorrgb#1{\color[rgb]{#1}}%
      \def\colorgray#1{\color[gray]{#1}}%
      \expandafter\def\csname LTw\endcsname{\color{white}}%
      \expandafter\def\csname LTb\endcsname{\color{black}}%
      \expandafter\def\csname LTa\endcsname{\color{black}}%
      \expandafter\def\csname LT0\endcsname{\color[rgb]{1,0,0}}%
      \expandafter\def\csname LT1\endcsname{\color[rgb]{0,1,0}}%
      \expandafter\def\csname LT2\endcsname{\color[rgb]{0,0,1}}%
      \expandafter\def\csname LT3\endcsname{\color[rgb]{1,0,1}}%
      \expandafter\def\csname LT4\endcsname{\color[rgb]{0,1,1}}%
      \expandafter\def\csname LT5\endcsname{\color[rgb]{1,1,0}}%
      \expandafter\def\csname LT6\endcsname{\color[rgb]{0,0,0}}%
      \expandafter\def\csname LT7\endcsname{\color[rgb]{1,0.3,0}}%
      \expandafter\def\csname LT8\endcsname{\color[rgb]{0.5,0.5,0.5}}%
    \else
      \def\colorrgb#1{\color{black}}%
      \def\colorgray#1{\color[gray]{#1}}%
      \expandafter\def\csname LTw\endcsname{\color{white}}%
      \expandafter\def\csname LTb\endcsname{\color{black}}%
      \expandafter\def\csname LTa\endcsname{\color{black}}%
      \expandafter\def\csname LT0\endcsname{\color{black}}%
      \expandafter\def\csname LT1\endcsname{\color{black}}%
      \expandafter\def\csname LT2\endcsname{\color{black}}%
      \expandafter\def\csname LT3\endcsname{\color{black}}%
      \expandafter\def\csname LT4\endcsname{\color{black}}%
      \expandafter\def\csname LT5\endcsname{\color{black}}%
      \expandafter\def\csname LT6\endcsname{\color{black}}%
      \expandafter\def\csname LT7\endcsname{\color{black}}%
      \expandafter\def\csname LT8\endcsname{\color{black}}%
    \fi
  \fi
    \setlength{\unitlength}{0.0500bp}%
    \ifx\gptboxheight\undefined%
      \newlength{\gptboxheight}%
      \newlength{\gptboxwidth}%
      \newsavebox{\gptboxtext}%
    \fi%
    \setlength{\fboxrule}{0.5pt}%
    \setlength{\fboxsep}{1pt}%
    \definecolor{tbcol}{rgb}{1,1,1}%
    \scalebox{0.6}{
\begin{picture}(7200.00,5040.00)%
    \gplgaddtomacro\gplbacktext{%
      \csname LTb\endcsname
      \put(682,1038){\makebox(0,0)[r]{\strut{}$1$}}%
      \put(682,1456){\makebox(0,0)[r]{\strut{}$2$}}%
      \put(682,1874){\makebox(0,0)[r]{\strut{}$3$}}%
      \put(682,2291){\makebox(0,0)[r]{\strut{}$4$}}%
      \put(682,2709){\makebox(0,0)[r]{\strut{}$5$}}%
      \put(682,3127){\makebox(0,0)[r]{\strut{}$6$}}%
      \put(682,3545){\makebox(0,0)[r]{\strut{}$7$}}%
      \put(682,3963){\makebox(0,0)[r]{\strut{}$8$}}%
      \put(682,4381){\makebox(0,0)[r]{\strut{}$9$}}%
      \put(682,4799){\makebox(0,0)[r]{\strut{}$10$}}%
      \put(1632,484){\makebox(0,0){\strut{}$5$}}%
      \put(2462,484){\makebox(0,0){\strut{}$10$}}%
      \put(3292,484){\makebox(0,0){\strut{}$15$}}%
      \put(4122,484){\makebox(0,0){\strut{}$20$}}%
      \put(4952,484){\makebox(0,0){\strut{}$25$}}%
      \put(5782,484){\makebox(0,0){\strut{}$30$}}%
    }%
    \gplgaddtomacro\gplfronttext{%
      \csname LTb\endcsname
      \put(209,2761){\rotatebox{-270}{\makebox(0,0){\strut{}$\nu$}}}%
      \put(3304,154){\makebox(0,0){\strut{}$x$}}%
      \csname LTb\endcsname
      \put(6300,704){\makebox(0,0)[l]{\strut{}$-10$}}%
      \put(6300,1732){\makebox(0,0)[l]{\strut{}$-5$}}%
      \put(6300,2761){\makebox(0,0)[l]{\strut{}$0$}}%
      \put(6300,3790){\makebox(0,0)[l]{\strut{}$5$}}%
      \put(6300,4819){\makebox(0,0)[l]{\strut{}$10$}}%
    }%
    \gplbacktext
    \put(0,0){\includegraphics[width={360.00bp},height={252.00bp}]{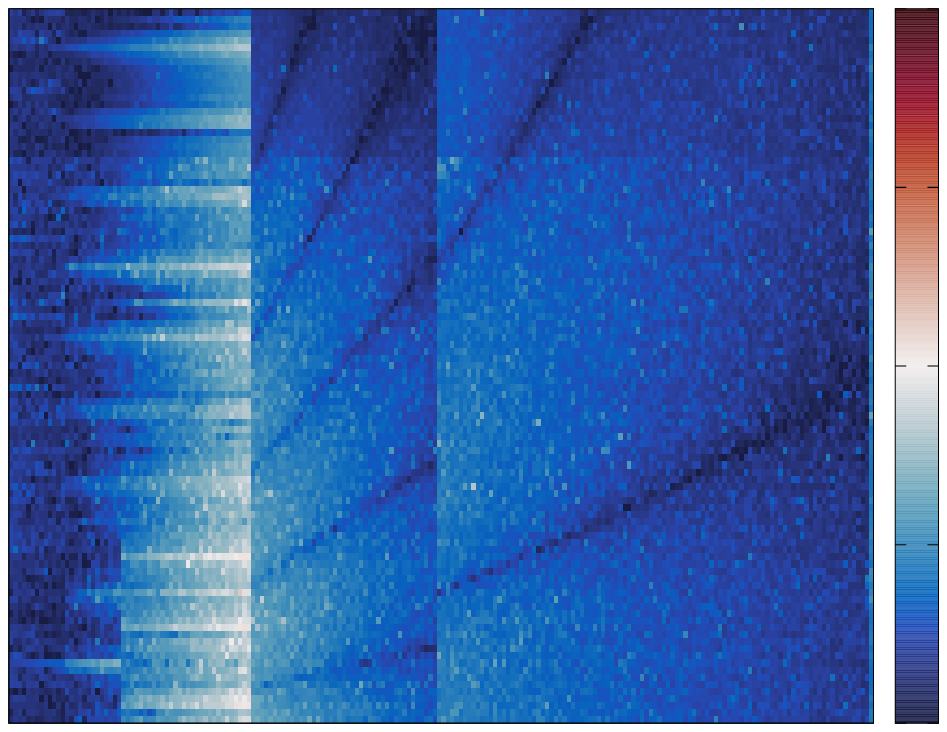}}%
    \gplfronttext
  \end{picture}%
  }
\endgroup
  \end{tabular}
  \caption{A log$_{10}$-scale comparison between
  $\partial_\nu^{k\text{AD}}\tilde{\besK}$ and $\partial_\nu^{k\text{FD}}\besK$
  using Equation \ref{eq:directlogdif}: first derivatives (left) and 
  second derivatives (right).}
  \label{fig:tols_log}
\end{figure}

To provide a relative comparison of derivative computations using finite
differences with \pkg{AMOS} versus \pkg{BesselK.jl}, we introduce the quantity
\begin{equation} \label{eq:directlogdif}
  \log_{10} \abs{ \partial_\nu^{k\text{R}} \besK(x) - \partial_\nu^{k\text{AD}} \tilde{\besK}(x) } - 
  \log_{10} \abs{ \partial_\nu^{k\text{R}} \besK(x) - \partial_\nu^{k\text{FD}} \besK(x) }
\end{equation}
for both first derivatives ($k=1$) and second derivatives ($k=2$).  By assessing
the error in log$_{10}$-scale, we can identify correct digits, and by taking the
difference between the errors of the two methods, we outline the regions where
AD outperforms finite differences.  In Figure~\ref{fig:tols_log} negative
numbers \revision{(blue hues)} indicate regions where $\partial_\nu^{\text{AD}}
\tilde{\besK}$ is more accurate than $\partial_{\nu}^{\text{FD}} \besK$ (with a
value of $-1$ indicating one additional digit of accuracy for the AD method),
and we note there are almost no positive values \revision{(red hues)} over the
entire domain. Upon close scrutiny we observe that the branch at $x\approx 8.5$
may incur losses of one to two digits in a few discrete locations (visible as
traces of red), which could be remedied by branching earlier than recommended at
$a_1<8.5$.  This is a fine-tuning aspect which vanishes for second derivatives
and was not explored extensively since it had no noticeable impact on our
results. Finite differences, notoriously inaccurate for second-order derivatives
since they involve a division by $h^2$, incur increasingly higher round-off
errors. In this context AD is expected to exhibit superior accuracy, and in
Figure~\ref{fig:tols_log} its advantage over finite differences is clearly
visible, consistently yielding $5$ or more extra digits of accuracy. 

\subsection{Efficiency diagnostics} 
In order to compare the efficiency of $\partial_\nu^{k\text{AD}} \tilde{\besK}$
with $\partial_\nu^{k\text{FD}} \besK$, we select pairs $(\nu, x)$ that cover
every code branch (outlined in Figure~\ref{fig:flowchart}) in the
implementation of $\besK$ and provide evaluation and derivative evaluation
timings performed on an Intel Core i$5$-11600K CPU in
Table~\ref{tab:speedresults}.  As can be seen, the speedup of $\tilde{\besK}$ is
substantial, including in regions of the domain in which its accuracy rivals
that of $\besK^A$, as well as for all derivatives.  The accuracy and efficiency
considerations discussed here are summarized in Table \ref{tab:test_summary} to
serve as a convenient reference to the reader.

\begin{table}[ht!]
  \centering
  \scalebox{0.75}{
    \begin{tabular}{|c|c|c|c|c|c|c|c|}
      \hline
      $(\nu, x) \rule{0pt}{2.6ex}$ 
      & $\tilde{\besK}$  
      & $\revision{\besK^A}$  
      & $\partial_\nu^{\text{AD}} \tilde{\besK}$  
      & $\partial_\nu^{\text{FD}} \besK$ 
      & $\partial_\nu^{2 \; \text{AD}} \tilde{\besK}$ 
      & $\partial_\nu^{2 \; \text{FD}} \besK$ 
      & case \\
      \hline
(0.500, 1) & 9 & 51 & 67 & 455 & 82 & 510 & half-integer (iv) \\
(1.000, 1) & 136 & 159 & 178 & 447 & 328 & 605 & whole integer (i) \\
(3.001, 1) & 139 & 285 & 181 & 570 & 337 & 847 & near-integer order (i) \\
(3.001, 8) & 227 & 300 & 284 & 597 & 557 & 892 & near-integer order, borderline arg (i) \\
(1.850, 1) & 85 & 229 & 139 & 460 & 307 & 686 & small order (ii) \\
(1.850, 8) & 103 & 241 & 293 & 483 & 531 & 722 & small order, borderline arg (ii) \\
(1.850, 14) & 167 & 209 & 281 & 416 & 568 & 624 & intermediate arg (iii) \\
(1.850, 29) & 94 & 191 & 149 & 380 & 293 & 565 & large intermediate arg (iii) \\
(1.850, 35) & 92 & 183 & 148 & 368 & 293 & 548 & large argument (iii) \\
      \hline
    \end{tabular}
  }
  \caption{Timings for evaluating $\besK^A$ and $\tilde{\besK}$ and their
  derivative methods (in units of ns) at various pairs $(\nu, x)$. 
The case column additionally provides references to the branch labels
  used in Figure \ref{fig:flowchart}.}
  \label{tab:speedresults}
\end{table}

\begin{table}[!ht]
\centering
\begin{tabular}{ |c|c|c| } 
 \hline
  Comparison & Higher accuracy & Higher efficiency  \\
 \hline
  $\besK^A$ \rule{0pt}{2.6ex} vs. $\tilde{\besK}$ 
 & 
 $\hphantom{\partial_\nu^\text{(F)}}\besK^A$ ($\approx$ 2-5 digits) 
 & 
 $\hphantom{\partial_\nu^\text{(A)}}\tilde{\besK}$ ($\approx$ factor of 1-3) \\
 $\partial_\nu^{\text{FD}} \besK$ vs. $\partial_\nu^{\text{AD}} \tilde{\besK}$ 
 & 
 $\partial_\nu^{\text{AD}} \tilde{\besK}$ ($\approx$ 2-5 digits) 
 &
 $\partial_\nu^{\text{AD}} \tilde{\besK}$ ($\approx$ factor of 2-3) \\
  $\partial_\nu^{2 \text{FD}} \besK$ vs. $\partial_\nu^{2 \text{AD}} \tilde{\besK}$ 
  &
 $\partial_\nu^{2 \text{AD}} \tilde{\besK}$ ($\approx$ $5+$ digits) 
 &
$\partial_\nu^{2 \text{AD}} \tilde{\besK}$ ($\approx$ factor of 2-5) \\
 \hline
\end{tabular}
\caption{A summary of the accuracy and speed diagnostics performed in this
  section. The accuracy comparison is given in correct digits, while efficiency
  is given as speedup factor.}
\label{tab:test_summary}
\end{table}

\subsection{Selected diagnostics for Mat\'ern covariance matrices}
\label{sec:covmats} The differences in accuracy and efficiency between the
\pkg{AMOS} library and \pkg{BesselK.jl} are most relevant in the context of
Mat\'ern covariance matrices. To this end we compare in Table
\ref{tab:maternresults} several key quantities for the Mat\'ern covariance
matrices $\bS$ and $\tilde{\bS}$, generated using $\besK^A$ and $\tilde{\besK}$
respectively. Points are chosen on a $24 \times 24$ grid on the domain
$[0,1]^2$, and pairs $(\rho,\nu)$ are chosen to cover various argument ranges
and branches in $\tilde{\besK}$ with respect to $\nu$, with both parameters
having significant effects on the condition of covariance matrices.

\begin{table}[!ht]
  \centering
  \scalebox{0.75}{
    \begin{tabular}{|c|c|c|c|c|c|c|c|c|}
      \hline
      $(\rho, \nu)$ & 
      $\tilde{T}_{\text{assembly}}$ &
      $T_{\text{assembly}}$ &
      $\tilde{\lambda}_{\text{min}}$ &
      $\lambda_{\text{min}}$ &
      $\abs{\lambda_{\text{dif}}}$ &
      $\log |\tilde{\bS}|$ &
      $\log \abs{\bS}$ &
      $|\log | \tilde{\bS} | - \log |\bS||$ \rule{0pt}{2.6ex}\\
      \hline
(0.010, 0.40) & 4.3e-02 & 4.3e-02 & 9.52e-01 & 9.52e-01 & 1.22e-12 & -2.60e-01 & -2.60e-01 & 3.11e-13 \\
(0.010, 1.25) & 4.4e-02 & 4.3e-02 & 9.79e-01 & 9.79e-01 & 3.47e-12 & -3.45e-02 & -3.45e-02 & 9.23e-12 \\
(0.010, 3.50) & 3.5e-02 & 3.5e-02 & 9.93e-01 & 9.93e-01 & 0.00e+00 & -3.14e-03 & -3.14e-03 & 0.00e+00 \\
(1.000, 0.40) & 2.8e-02 & 4.8e-02 & 3.78e-02 & 3.78e-02 & 9.76e-15 & -1.40e+03 & -1.40e+03 & 1.84e-11 \\
(1.000, 1.25) & 2.6e-02 & 5.4e-02 & 1.03e-04 & 1.03e-04 & 6.56e-14 & -4.04e+03 & -4.04e+03 & 1.19e-09 \\
(1.000, 3.50) & 2.4e-02 & 3.5e-02 & 7.18e-11 & 7.18e-11 & 1.02e-14 & -1.02e+04 & -1.02e+04 & 2.76e-03 \\
(100.000, 0.40) & 2.7e-02 & 4.5e-02 & 9.50e-04 & 9.50e-04 & 1.15e-14 & -3.51e+03 & -3.51e+03 & 6.96e-10 \\
(100.000, 1.25) & 2.5e-02 & 4.6e-02 & 1.03e-09 & 1.03e-09 & 2.03e-15 & -1.06e+04 & -1.06e+04 & 3.87e-05 \\
(100.000, 3.50) & 1.7e-02 & 3.8e-02 & -3.43e-13 & -2.72e-13 & 7.15e-14 & NaN & NaN & NaN \\
      \hline
    \end{tabular}
  }
  \caption{Summary properties of $\bS$ and $\tilde{\bS}$ for various pairs
  $(\rho, \nu)$. Quantities with tildes, such as $\tilde{\lambda}$, are derived
  from $\tilde{\bS}$, which has been built using $\tilde{\besK}$. Quantities
  with no tildes were derived from $\bS$, which was built with $\besK^A$. In the
  last row, values of \texttt{NaN} are a result of the Cholesky factorization
  failing, which occurred for both implementations.}
  \label{tab:maternresults}
\end{table}

As expected from the previous benchmarks, the matrix assembly speed,
$\tilde{T}_{\text{assembly}}$ for \pkg{BesselK.jl}, is superior to \pkg{AMOS}
\revision{in regions where the predominant evaluation strategies for $\besK$ is
different for each library (thus excluding, for example, cases like $\rho=0.01$
in which both software libraries are using asymptotic expansions)} and
approximately equal otherwise.  As a preliminary investigation of matrix
similarity, we compare the smallest eigenvalues of each matrix,
$\tilde{\lambda}_{\text{min}}$ and $\lambda_{\text{min}}$, and note that they
agree to almost computer precision in all cases. We also compare
log-determinants, as they are directly used in maximum likelihood estimation,
and in most cases the log-determinants agree to high precision, although two
situations stand out as being slightly inaccurate compared to the others. These
two cases have a combination of sufficiently large $\rho$ and $\nu$ parameters,
yielding especially ill-conditioned matrices that pose numerical challenges even
when $\besK$ is computed to double precision accuracy. With this in mind, we
consider the numerical agreement in these two cases to be satisfactory.

\section{A Demonstration with Simulated Data}\label{sec:demo}
 
In this section, we demonstrate that in some cases even the slightest numerical
inaccuracy in finite difference derivatives can accumulate so catastrophically
as to yield completely incorrect Hessian matrices of the Gaussian
log-likelihood, which can in turn lead to failure in second-order maximum
likelihood estimation.  

To demonstrate this we select $512$ random locations on the unit square
$[0,1]^2$ and simulate ten independent realizations from a Mat\'ern Gaussian
process with parameters $(\sigma, \rho, \nu) = (1.5, 2.5, 1.3)$. Subsequently,
we perform maximum likelihood estimation with the \pkg{Ipopt} library
\citep{wachter2006} using manually computed gradients and Hessians via the
formulae provided in the introduction. In all cases, for derivatives involving
only $\sigma$ and $\rho$, we use analytical derivatives of the Mat\'ern
covariance function. To study the effect of using finite difference derivatives
versus our proposed AD derivatives, we consider \revision{six} different
optimization problems. First, we compute the first two derivatives of
$\mathcal{M}_\nu(\bx, \bx')$ with respect to $\nu$ using finite difference
approximations to compute Hessian matrices. In the second setting, we compute
gradient and Hessian information using AD-generated derivatives and again
optimize with true Hessians.  In the next two settings, we again use finite
differences and AD for derivatives that pertain to $\nu$, but instead of
Hessians we use expected Fisher matrices as proxies.  \revision{Recalling that
$\bS_j(\bth) = \tfrac{\partial}{\partial \theta_j} \bS(\bth)$,} the expected
Fisher information, given by
$
  \mathbf{I}_{j,k} := \frac{1}{2} \text{tr} \left(
    \bS(\bth)^{-1} \bS_j(\bth)
    \bS(\bth)^{-1} \bS_k(\bth)
  \right),
$
is, under appropriate regularity conditions, the asymptotic precision of maximum
likelihood estimators. Unfortunately, these conditions are not often met for
spatial processes under fixed-domain asymptotics \citep{stein1999}; nonetheless
it is a valuable proxy for Hessian information. Moreover, it only requires first
derivatives of the covariance function to compute, providing a natural way to
separately investigate the effect of first and second finite difference
derivatives.  \revision{The final two settings are in a similar spirit and again
investigate using only first derivative information, but now using the more
general-purpose BFGS approximation for Hessian information.}

While the parameters for the simulated data have been chosen purposefully, they
are not implausible for much environmental data. The range parameter $\rho =
2.5$ is large compared to the domain radius, which, even with 10 replicates,
makes estimating the range difficult \citep{zhang2004}. This level of
dependence could occur in, for example, daily values of meteorological
quantities such as temperature or pressure over a region of diameter on the
order of 100 km.  A smoothness of $\nu = 1.3$ is likewise a very plausible value
for a quantity like atmospheric pressure, which should be smooth but not too
smooth. The primary challenge with these parameter values is that they require
derivatives of $\besK$ very near the origin, which we have demonstrated in the
previous section to be particularly problematic. With this said, 
computing the upper $2 \times 2$ block of the Hessian with analytical
derivatives in both cases limits the scope of impact of finite difference
approximations as much as possible.  Using finite difference derivatives
directly on $\mathcal{M}_\nu$, which is well-behaved when its argument is near
0, rather than on $\besK$ directly, should also be favorable to finite
differences.

\begin{table}[!ht]
  \centering
  \begin{tabular}{|c|c|c|c|c|}
    \hline
    method & convergence & \#iterations & terminal log-likelihood & MLE$^*$ \\
    \hline
    FD (BFGS)    & No  & $100$ & $-19192.36$ & $(1.574,  2.642,  1.060)^*$ \\
    FD (Fisher)  & No  & $100$ & $-21548.71$ & $(1.576,  2.642,  1.293)^*$ \\
    FD (Hessian) & No  & $100$ & $-21473.38$ & $(47.428, 24.628, 1.380)^*$ \\
    AD (BFGS)    & No  & $100$ & $-21548.71$ & $(1.576,  2.642,  1.293)^*$ \\
    AD (Fisher)  & Yes & $58$  & $-21548.71$ & $(1.576,  2.642,  1.293)$   \\
    AD (Hessian) & Yes & $25$  & $-21548.71$ & $(1.576,  2.642,  1.293)$   \\
    \hline
  \end{tabular}
  \caption{A comparison of optimization results using second-order optimization
  with finite difference derivatives of the Mat\'ern covariance versus automatic
  differentiation of the Mat\'ern covariance with \pkg{BesselK.jl}. In the MLE
  column, asterisks indicate terminal return parameters in the case of failed
  optimization by reaching the maximum allowed number of iterations, here chosen
  to be $100$.}
  \label{tab:optimization}
\end{table}

Table \ref{tab:optimization} summarizes the results of the estimation
procedures. In the case of true second-order optimization with finite difference
derivatives used for Hessians, the optimization completely fails.  Optimization
using the expected Fisher matrix built with finite difference derivatives does
successfully reach the MLE parameters to high accuracy, but due to slight
inaccuracies in derivatives fails to reach the termination criteria tests used
by \pkg{Ipopt}. Using AD derivatives to assemble the expected Fisher matrix
solves this problem, and the optimization successfully terminates after $58$
iterations. Finally, optimization with the Hessians built with AD-generated
derivatives is fast, accurate, and terminates successfully in just $25$
iterations. These results are consistent across various starting values and
termination criterion tolerances.

To better understand why the second-order optimization with FD-based Hessians
fails, we inspect these Hessians at several values. Starting at
the initializer, we look at Hessian matrices computed using the two different
sources of input derivatives. As a point of reference, we also provide a very
high-order adaptive finite difference Hessian using the log-likelihood function
itself. As above, the parameter order here is $(\sigma, \rho, \nu)$, so that the
third column and row are the entries that pertain to derivatives of
$\mathcal{M}_\nu$ with respect to $\nu$. Looking at the generic initializer of
all parameters being equal to $1$, we inspect the third column, using the
shorthand $\bm{1}_3$ for $(\sigma, \rho, \nu) = (1,1,1)$:
\begin{align*} 
  \left[ H^{\text{ref}} \ell_{\bm{1}_3} \right]_{\cdot, 3} 
  = 
  \matr{
    -1104.05 \\
    3712.60  \\
    5173.03 
  }
  \hspace{0.1in}
  \left[ H^{\text{AD}}  \ell_{\bm{1}_3} \right]_{\cdot, 3} 
  = 
  \matr{
    -1104.05 \\
    3712.60  \\
    5173.03 
  }
  \hspace{0.1in}
  \left[ H^{\text{FD}}  \ell_{\bm{1}_3} \right]_{\cdot, 3} 
  = 
  \matr{
    -1100.47 \\
    3716.86  \\
    4533.61
  } \ .
\end{align*}
What immediately stands out in this comparison is that the component pertaining
to the second derivative $\partial_\nu^2 \besK$ with finite difference-built
Hessians is grossly inaccurate. For inaccuracies at this level, one has reason
to be concerned that very important quantities for optimization, for example the
unconstrained Newton direction $-H \ell_{\bth}^{-1} \nabla \ell_{\bth}$, will be
materially affected in a way that inhibits the progress of an optimizer.

Looking now at the third columns of these three matrices at the MLE, we see this
problematic inaccuracy progressing further, using $\hat{\bth}$ as shorthand for
the MLE:
\begin{align*} 
  \left[ H^{\text{ref}} \ell_{\hat{\bth}} \right]_{\cdot, 3} 
  = 
  \matr{
    -23920.00 \\
     18431.04 \\
     145597.78
  }
  \hspace{0.1in}
  \left[ H^{\text{AD}}  \ell_{\hat{\bth}} \right]_{\cdot, 3} 
  = 
  \matr{
    -23920.01 \\
     18431.04 \\
     145597.79
  }
  \hspace{0.1in}
  \left[ H^{\text{FD}}  \ell_{\hat{\bth}} \right]_{\cdot, 3} 
  = 
  \matr{
    -23919.54 \\
     18430.66 \\
     135175.16
  } \ .
\end{align*}
We observe once again that the AD-generated Hessians are very accurate.  The
FD-based Hessian is reasonably accurate save for the component pertaining to
$\partial_\nu^2 \besK$. But as before, this quantity is sufficiently inaccurate
that it necessarily will change important optimization quantities.  Moreover,
the matrix $H^{\text{FD}} \ell_{\hat{\bth}}$ is not even close to being positive
semi-definite (with a minimum eigenvalue of $-192.1$ compared to $H^{\text{AD}}
\ell_{\hat{\bth}}$'s minimum eigenvalue of $4.036$). Observing FD-generated
Hessian matrices as the range parameter continues to increase, one will see a
continual worsening of this specific component, which is arguably a particularly
serious issue because it introduces structured inaccuracy to the optimization
problem. This inaccuracy's effect on derived quantities for optimization such as
search directions is a very likely explanation for the complete failure in the
optimization using \pkg{Ipopt}, which unlike simpler optimization software will
pursue candidate directions in search of a feasible minimizer even if they do
not immediately decrease the objective function value.  

The conclusion from these results is clear: second-order optimization using
finite difference methods for second derivatives of $\mathcal{M}_\nu$, and by
extension $\besK$, presents a significant risk. As a final observation, we reiterate
 that the demonstration chosen here is the best possible case for finite
differences in that only the third row and column of these Hessians involve any
approximation, and only one element of the Hessian matrix was incorrect to a
degree of obvious concern. Even with the numerical issues of finite differences
restricted to this small setting, the optimization results were a complete
failure.  For even slightly more complicated parametric models, such as adding a
geometric anisotropy, the number of Hessian entries that involve finite
difference derivatives will be greater, further increasing the risk of
accumulated inaccuracies and severely incorrect point estimates.

\section{Summary and Discussion}

In this work, we identified a set of numerical approaches whose regions of
accuracy and efficiency cover the entire domain of $\besK$ and that are
well-suited to high-order automatic differentiation. After assessing the computational
gains for first and second derivatives our approach provides, we demonstrated
the practical significance of the accuracy gains through an example of maximum
likelihood estimation via second-order optimization that completely fails when
finite difference derivatives are employed, despite the finite difference
derivatives being computed with the highly accurate \pkg{AMOS} library.

As modern datasets have continued to grow in size and complexity, accurately
modeling dependence structure is more important than ever, and Gaussian
processes have continued to serve as among the most popular tools for that
purpose. We consider the full three-parameter isotropic Mat\'ern model, in which
one estimates $\nu$, to be a bare minimum level of flexibility, at least for
many datasets in the physical sciences. We thus hope that this work will empower
practitioners to fit $\nu$ as easily as they do any other parameter. In order to
make this as straightforward as possible, we provide the methods described in
this paper in a convenient and freely available software package.
\revision{While this work provides no discussion of scalable approximations to
Gaussian likelihoods, they are a popular and important area of contemporary
research in the field (see \citep{heaton2019}, for example, for a non-exhaustive
but nonetheless expansive introduction to the topic). Particularly in the
fixed-domain asymptotic regime, large datasets are especially informative about
the smoothness parameter $\nu$, and so there is particular synergy between the
kernel derivatives we discuss here and scalable likelihood approximations that
can use the information of significantly more data than would traditionally be
possible with direct dense linear algebra}.

Additionally, we hope that this work will reduce the complexity barrier to using
Hessian matrices of the log-likelihood. Beyond the demonstration that we provide
here of the value of true second-order optimization for maximum likelihood
estimation, Hessian matrices of the log-likelihood are crucial to other common
methods, such as Laplace approximations, including the method of integrated
nested Laplace approximations (INLA) \citep{Rue2009}. It is thus our hope that
this work will be helpful to a wide variety of scientists and will make the use
of the full three-parameter Mat\'ern covariance a more standard practice.

\section*{Acknowledgements}
The authors would like to thank Paul Hovland (Argonne National Laboratory) for
engaging discussions on automatic differentiation as well as suggestions of
relevant literature on the topic. They are also grateful to Lydia Zoells for her
careful copyediting. \revision{Finally, they are very grateful to the two
anonymous referees, both of whom provided exceptionally helpful comments and
suggestions.}

\vspace*{1em}
\noindent\fbox{\parbox{\columnwidth}{\footnotesize
    The submitted manuscript has been created by UChicago Argonne, LLC, Operator
    of Argonne National Laboratory (``Argonne''). Argonne, a U.S. Department of
    Energy Office of Science laboratory, is operated under Contract No.
    DE-AC02-06CH11357. The U.S. Government retains for itself, and others acting
    on its behalf, a paid-up nonexclusive, irrevocable worldwide license in said
    article to reproduce, prepare derivative works, distribute copies to the
    public, and perform publicly and display publicly, by or on behalf of the
    Government. The Department of Energy will provide public access to these
    results of federally sponsored research in accordance with the DOE Public
    Access Plan (http://energy.gov/downloads/doe-public-access-plan).}
}

\begin{appendix}

\section{Modifications for rescaling} \label{sec:rescale}

In certain cases a \emph{scaled} modified Bessel function, $e^x \besK(x)$, may
be useful in extending the interval of definition to larger arguments and
avoiding numerical issues such as underflow. In the Mat\'ern covariance
function, a different prefactor comes up: $x^\nu \besK(x)$. This function, while
more sensitive to floating point round-off errors than its unscaled counterpart,
is bounded at the origin for all orders $\nu > 0$. 

This rescaled modified Bessel function is primarily useful for small arguments,
and the two routines that can benefit the most from this rescaling are the
direct series from Section~\ref{sec:besk_ser} and the Temme recursion in
Section~\ref{sec:int_orders}.  In Equation~\ref{eq:besk_ser}, one can simply
bring the $x^\nu$ term inside the sum to cancel the $x^{-\nu} x^\nu$ product,
completely removing the singularity at the origin as well as lowering the
operation count, for both direct evaluations and AD computations.

To modify the expressions in Section~\ref{sec:int_orders} is slightly more
complicated, since integer values of $\nu$ are computed via a recursion starting
from the $\besK[0]$ and $\besK[1]$ terms.  However, at $\nu=0$ the term
$x^0\besK[0]$ contains a (logarithmically growing) singularity, problematic also
for integer orders $\nu$ when $x=0$. We have observed this can be avoided since
the recursion can be re-arranged to be
\begin{equation*} 
  x^{\nu+1} \besK[\nu+1](x) = 2 \nu (x^\nu \besK(x)) - x^2 (x^{\nu-1}
  \besK[\nu-1](x)) ,
\end{equation*}
where for $x=0$ the term starting off $\besK[0]$, i.e.  $\besK[\nu-1]$, cancels
entirely, simplifying the recursion for all integer orders. Introducing a branch
in the code that checks for this condition results in AD-compatible
differentiations of $x^\nu \besK(x)$ for non-zero integer values $\nu$.

When $\nu \not\in \Z$ and $x=0$, as the terms $x^\nu \sinh(\mu)/\mu$ and $x^\nu
\cosh(\mu)$ cannot be evaluated directly, and we have a secondary set of limit
branch problems. In this case we observed that $x^\nu \cosh(\nu \log(2/x))$ can
be re-written as
\begin{equation*} 
  x^\nu \cosh(\nu \log(2/x)) =
  \frac{x^\nu (2^\nu x^{-\nu} + 2^{-\nu} x^\nu)}{2}\ .
\end{equation*}
By again manually performing the cancellations, we can preserve the derivative
information for $x^\nu \cosh(\mu)$, and similarly for $\sinh(\mu)$. This minor
accountancy addition, which introduces special routines for $x^\nu \besK(x)$ for
all orders $\nu$, pays off in terms of accuracy, speed, and AD-simplicity.

\section{Profile likelihoods} \label{sec:appendix_profile}

\revision{
For a collection of parameters $\bth = (\sigma^2, \rho, \nu, ...)$, let $\bth_1$
denote the collection $(1, \rho, \nu, ...)$. Through a small amount of algebra,
it can be observed that, given all parameters besides $\sigma^2$, the value of
$\sigma^2$ that maximizes the log-likelihood can be computed in closed-form as
$n^{-1} \bz^T \bS(\bth_1) \bz$. Due to the observation that $\bS(\bth) =
\sigma^2 \bS(\bth_1)$ and some subsequent algebraic manipulations, one can thus
re-write the likelihood $\ell(\bth)$ as a function of all parameters
\emph{besides} $\sigma^2$ to obtain 
\begin{equation*} 
  -2 \ell_p(\bth_1 \sv \bz) = 
  \log \abs{\bS(\bth_1)} + n \log \left(
    \bz^T \bS(\bth_1)^{-1} \bz
  \right).
\end{equation*}
The value of this reformulation is that the dimension of the parameter
estimation problem (and thus the challenging nonconvex optimization problem) has
been reduced by one. For this paper, however, we only utilize the profile
log-likelihood in order to visualize likelihood surfaces that would otherwise
be three-dimensional.
}

\end{appendix}
\bibliography{spfun_short}
\end{document}